\documentclass[english]{iopart}

\usepackage{graphicx}
\usepackage{url}
\usepackage{iopams} 

\expandafter\let\csname equation*\endcsname\relax
\expandafter\let\csname endequation*\endcsname\relax

\usepackage{amssymb}
\usepackage[latin1]{inputenc}
\usepackage{amsmath}
\usepackage{epsfig}
\usepackage{amsfonts}

\usepackage{url}
 \usepackage{hyperref}
 \hypersetup{pdfstartview={XYZ null null 1.00},pdfpagemode=UseNone}

\newcommand{\beq}{\begin{equation}}
\newcommand{\eeq}{\end{equation}}

\newcommand{\beqr}{\begin{eqnarray}}
\newcommand{\eeqr}{\end{eqnarray}}

\newcounter{fig}
\begin{document}

\title[Perimeter g.f. of 3-choice, imperfect, 1-punctured staircase polygons]
{\Large The perimeter generating functions of three-choice, imperfect, 
and 1-punctured staircase polygons}

\vskip .2cm 

\author{M. Assis$^1$, M. van Hoeij$^2$ and J-M. Maillard$^3$}
\address{$^1$ MASCOS, School of Mathematics and Statistics, 
University of Melbourne, Carlton, VIC, Australia}
\address{$^2$ Florida State University, Department of Mathematics, 
1017 Academic Way, Tallahassee, FL 32306-4510 USA}
\address{$^3$ LPTMC, UMR 7600 CNRS, Universit{\'e} de Paris 6, 
Tour 23, 5{\`e}me  {\'e}tage, 
case 121, 4 Place Jussieu, 75252 Paris Cedex 05, France}
\ead{michael.assis@unimelb.edu.au, hoeij@mail.math.fsu.edu, maillard@lptmc.jussieu.fr}

\vskip .2cm 

\vskip .2cm 

{\em Dedicated to A. J. Guttmann on the occasion of his 70th birthday}

\vskip .2cm

\begin{abstract}

We consider the isotropic perimeter generating functions of three-choice, 
imperfect, and 1-punctured staircase polygons, whose 8th order linear Fuchsian 
ODEs are previously known. We derive simple relationships between the 
three generating functions, and  show that all three generating functions 
are joint solutions of a common 12th order Fuchsian linear ODE.
We find that the 8th order differential operators 
can each be rewritten as a direct sum of a direct product, with operators no larger 
than 3rd order. We give closed-form expressions for all the solutions of
these operators in terms of ${}_2F_1$ hypergeometric functions with 
rational and algebraic 
arguments. The solutions of these linear differential operators
can in fact be expressed in terms of two modular forms, since these 
${}_2F_1$ hypergeometric functions can be expressed with two, rational or
algebraic, pullbacks. 

\end{abstract}


\vskip .1cm 

\noindent {\bf AMS Classification scheme numbers}: 03D05, 11Yxx, 33Cxx,  34Lxx, 
 34Mxx, 34M55,  39-04, 68Q70   

\vskip .3cm 


\vskip .3cm 

\section{Introduction}
\label{intro}

Self-avoiding walks (SAWs) and self-avoiding polygons (SAPs) have long been studied 
in enumerative combinatorics as models of percolation, polymers, surface roughness, 
and more~\cite{guttmann20092}, although both their generating functions remain 
unsolved to this day. Several classes of SAWs and SAPs have been solved by imposing 
either convexity or directedness constraints, or both. Within a class, walks 
and polygons usually have the same growth constant, also known as the connective 
constant, although recently prudent polygons have been shown to be exponentially 
sparse among prudent walks~\cite{garoni2009gjd}.

The study of SAPs and its sub-categories involves the search 
for exact expressions of their generating functions 
as a function of various parameters of interest. These include the perimeter, 
width, height, site perimeter, left and right corners, and area, and for certain 
classes of SAPs, a generating function has been found which include all of these 
parameters explicitly, e.g.~\cite{delest1995dd}. Among the known generating 
functions, rational, algebraic, $D$-finite, non $D$-finite, and natural boundaries have 
been derived  (see~\cite{rechnitzer2000} for a good review). Furthermore, among 
still unsolved classes, it is possible to prove results concerning the nature 
of the unsolved generating function. For example, the anisotropic perimeter 
generating function for the full SAPs class has been proven to not be a $\, D$-finite 
function in~\cite{rechnitzer2006}. The wide variety of types of functions which 
arise in the study of SAPs offers an intriguing source of knowledge for what 
constitutes exact solutions in statistical mechanics. 

Among known perimeter generating functions are rational functions, algebraic 
functions, $q$-series, and natural boundaries~\cite{rechnitzer2000}, and quite 
generically the nature of the isotropic and anisotropic perimeter 
generating functions are the same. In the case of column-convex 
but not row-convex SAPs, the area generating functions 
are simpler than the corresponding perimeter generating function, being 
rational functions~\cite{rechnitzer2000}. However, all known cases of area-perimeter 
generating functions involve q-series~\cite{rechnitzer2000}. 

Three-choice and 1-punctured staircase polygons are two classes of SAPs well studied 
in the literature~\cite{guttmann1993po,bousquetmelou1999gor}, known to be $\, D$-finite  
functions~\cite{guttmann2006j,guttmann2006j2} but whose perimeter generating functions 
have resisted closed-form solutions~\cite{guttmann20093}. We here provide
hypergeometric solutions to the operators appearing in their linear ODEs. 
It has long been suspected that their generating functions are related 
to each other~\cite{bousquetmelou1999gor,guttmann2006j,guttmann2006j2,guttmann2000jwe}, 
and indeed we here show that they are equal up to the sum of an algebraic 
factor. Our hypergeometric 
solutions constitute the first example of a SAP generating function which 
is $\, D$-finite but not algebraic. 

We begin by reviewing the literature of staircase, three-choice, and punctured staircase 
polygons in section~\ref{sec:known}, followed by an analysis of the linear differential 
operators of the three-choice and punctured staircase polygon linear ODEs in 
section~\ref{sec:ODE}. We then provide solutions for the linear differential operators 
in section~\ref{sec:results} and explore hypergeometric and modular function
identities of the solutions in section~\ref{sec:identities}. We end with a discussion 
of generalizations of the results in section~\ref{sec:discussion}, followed 
by conclusions in section~\ref{conclusions}.

\vskip .1cm 

\section{Known results}
\label{sec:known}

\subsection{Staircase polygons}
Staircase polygons are polygons formed from two self-avoiding walks that both start 
at the origin, move using only north or east steps (sometimes south and east 
steps~\cite{guttmann2000jwe}) and only intersect once again at their common 
endpoint. Even though they have a long history in enumerative combinatorics and are 
among the most well studied classes of SAPs, their literature can be difficult 
to navigate, with numerous erroneous references, many independent proofs, and 
multiple equivalent names. Viewed in terms of their area, they are often called 
parallelogram polyominoes. In~\cite{delest1993f} they are also called skew Ferrers 
diagrams, defined as the difference between two Ferrers diagrams. While it is
unstated in~\cite{delest1993f}, it is clear from~\cite{delest1990f} that only 
the connected skew Ferrers diagrams are being considered in~\cite{delest1993f}, 
such that indeed they correspond to staircase polygons. Finally, viewed in terms 
of two vicious walkers which start at the origin and end at their only other 
common point, they have also been called two-chain watermelons, or 
2-watermelons in~\cite{chen2011dz}.

Staircase polygons are examples of convex and directed SAPs and all typical 
quantities of interest are known exactly for them. Jack Levine appears to be 
the first to have published a proof of the isotropic and anisotropic perimeter 
generating functions in 1959~\cite{levine1959}. Nevertheless, his paper seems 
to have been largely neglected in the literature. P{\'o}lya in 1969 published 
the formula for the isotropic perimeter generating function, stated without 
proof but with reference to a diary entry from 1938~\cite{polya1969}. Other 
independent proofs have appeared, in 1984~\cite{delest1984v} and 
in 1987~\cite{lin1987mkc} for the isotropic perimeter. The site perimeter is 
a relevant quantity in percolation theory, and for staircase polygons it can be 
computed from the perimeter and the number of corners the polygon has. The 
perimeter-corner generating function is given in~\cite{delest1987gv}.

The inclusion of area in the generating function began with P{\'o}lya 
in 1969~\cite{polya1969}, who provided an expression for the area-perimeter 
generating function, stated without proof. An expression for the area generating 
function was first proven in 1974~\cite{klarner1974r}, followed by various proofs 
of the area-perimeter generating function as a continued 
fraction~\cite{gessel1980,bousquetmelou1992v}, as well as a ratio of $\, q$-Bessel 
functions~\cite{brak1990g,bousquetmelou1992v}. The area-width generating function 
was given in~\cite{delest1993f}, while the area-perimeter-left/right height 
generating function was given in~\cite{bousquetmelou1996}. The most general 
generating function, enumerated by area, perimeter, width, height, and left/right 
corners, was given in~\cite{delest1995dd}.

Here we collect a few expressions. The isotropic perimeter generating function of 
half-perimeter $n$ is related to the Catalan numbers $ \, C_n$
\beqr
\hspace{-0.95in}   \quad  \quad   \quad   \quad \quad 
P^{\mathrm{S}}\,\,  &=& 
\, \,  \, \sum_{n=2}^{\infty}\,  \binom{2n}{n} \cdot \,\frac{x^n}{(4n-2)}
 \,\,  \, = \,\, \, \,  \sum_{n=1}^{\infty}\,  C_n \cdot \, x^{n+1}
\nonumber \\ 
\label{Px}
\hspace{-0.95in}   \quad  \quad \quad   \quad \quad 
 \,\, &=& \,\, \,\,
\frac{1-2x\, -\sqrt{1-4x}}{2}\,\,  \, = \,\, \, 
 \frac{4x^2}{(1+\sqrt{1-4x})^2}. 
\eeqr
We note the single square root singularity at $x=1/4$.

The anisotropic perimeter generating function, in terms of $\, h$ horizontal 
and $\, v$ vertical perimeter, is given by 
\beqr
\label{Pxy}
\hspace{-0.95in}   \quad  \quad \quad 
P(x,y) \,\,&=& \,\,\,
 \sum_{h,v\geq 1}\binom{h+v-1}{h} 
  \binom{h+v-1}{v} \cdot \,  \frac{x^h y^v}{(h+v-1)} 
\\
\hspace{-0.95in}   \quad  \quad \quad \quad \quad \quad 
&=& \,\,\, \, \,\,  \frac{1}{2} \cdot
 \,\left(1\,-x-y\,\, \, -\sqrt{1\,-2x-2y\,\, +x^2+y^2-2xy} \right), 
\eeqr
where the previous result is obtained by setting 
$ \, y= \, x \to \sqrt{x}$. The anisotropic 
perimeter generating function satisfies the simple algebraic equation
\beqr
\label{PPxy}
\hspace{-0.95in}&&   \quad  \quad \quad  \quad \quad \quad \quad 
P \,\, \,  = \,\,  \, \, (P\, +x) \cdot \, (P\, +y), 
\eeqr
and the inversion relation
\beqr
\hspace{-0.85in}&&   \quad  \quad \quad \quad \quad 
P(x,y)\,\,\,  - x \cdot \, P\left(\frac{1}{x},\frac{y}{x} \right)
\, \,\,  = \, \, \,\, 1\, -x. 
\eeqr
Finally, we note the following functional equation for the isotropic 
half-perimeter $\, x$ and area $q$ generating function~\cite{richard2006}
\beqr
\hspace{-0.95in}&&   \quad  \quad \quad \quad \quad \quad 
P(x,q)\,\,\,   = \,\,\, \, 
\frac{q \cdot \, x^2}{1\,-2 \, q \cdot \, x \, -P(q \,x, \, q)}, 
\eeqr

\subsection{Three-choice and imperfect staircase polygons}
\label{Threechoicepol}

Three-choice SAWs were defined by Manna in 1984~\cite{manna1984} as SAWs 
where right-handed turns are disallowed after travelling in the 
east or west directions, and in 1993 their SAP equivalents 
were considered~\cite{guttmann1993po}. There are two classes 
of three-choice polygons, the usual staircase polygons and imperfect 
staircase polygons. Depending on the authors, ``three-choice polygons" 
can either mean both classes, or only the imperfect staircase polygons, 
e.g.~\cite{bousquetmelou1999gor}. 
A polynomial time algorithm for the enumeration of three-choice 
polygons by isotropic perimeter was given in~\cite{conway1997gd}, 
which hinted at its solvability. In that same work, it was shown using 
the theory of algebraic languages, that the perimeter generating 
function is not algebraic. Nevertheless, in~\cite{guttmann2006j2} its 
8th order linear ODE was found from a long series expansion, so that 
it is a $\, D$-finite transcendental function. The singularity closest 
to the origin on the positive real axis is at $x= \, 1/4$, the same 
location as for staircase polygons. 

From the analysis 
of its series expansion~\cite{rechnitzer2003}, 
it is also expected that the anisotropic perimeter generating 
function is both solvable and $\, D$-finite (see page 85 
of~\cite{guttmann20093} 
for the mention, see ref. [4] of~\cite{guttmann20093},
 of an unpublished proof
that it is $\, D$-finite). Furthermore, 
the generating function for the area and anisotropic 
perimeter was shown in~\cite{bousquetmelou1999gor} to satisfy 
self-reciprocity and inversion relations. The anisotropic 
generating function for imperfect staircase polygons satisfies 
the following inversion relation~\cite{guttmann1998ro}
\beqr
\hspace{-0.95in}  \quad  \quad  \quad  \quad  
&&P^{\mathrm{I}}(x,y) 
\, \, \, 
+ x^2 \cdot \, P^{\mathrm{I}}\left(\frac{1}{x},\frac{y}{x} \right)
 \nonumber\\
\hspace{-0.95in}  
&&\quad \quad   = \, \,  \,\, 
\frac{x^2-1}{2} \cdot \, 
\left(\frac{1-2x^2-y^2-x^2y^2+x^4 \, +(x^2-1)
 \cdot \, \sqrt{\Delta}}{\sqrt{\Delta}} \right),
\eeqr
where
\beqr
\hspace{-0.95in}  \quad  \quad  \quad   \quad   \quad   \quad  
\Delta \, \, = \, \, \, \, (1+x+y)\, (1+x-y)\, (1-x+y)\, (1-x-y).
\eeqr

The full three-choice polygon perimeter generating function series reads
\beq
\label{series1}
P^{\mathrm{T}}  \,= \, \,  \, \, 
4\, x^2 \,\,  +\, 12 \, x^3  \, +\, 42 \, x^4  \, 
+\,152 \, x^5 \, +\,562 \, x^6  +\, 2108\,x^7\, +\, 7986\,x^8\, 
\,\, + \,\, \, \cdots  
\eeq
and the subset of only imperfect staircase polygon perimeter generating 
function series reads
\beq
\label{series11}
P^{\mathrm{I}}  \,= \, \, \, \, 
x^4 \, \, +\, 6 \, x^5  \, +\, 29 \, x^6  \, 
+\,130 \, x^7 \, +\,561 \, x^8   +\, 2368\,x^9\,   +\, 9855\,x^{10}\,
\,\, + \,\, \, \cdots 
\eeq

\subsection{Punctured staircase polygons}
\label{Puc}

Punctured staircase polygons are staircase polygons with holes in the 
shape of a smaller staircase polygons whose perimeter does not share 
any vertices with the outer perimeter. We here only consider 1-punctured 
staircase polygons and below use ``punctured staircase polygons" synonymously 
with 1-punctured staircase polygons. Punctured staircase polygons were first 
considered in~\cite{bousquetmelou1999gor}, where the generating function for 
the area and anisotropic total perimeter were shown to satisfy 
self-reciprocity and inversion relations. In~\cite{guttmann2000jwe}, 
a polynomial time algorithm was given for the enumeration of the total 
isotropic perimeter generating function for one to three holes, hinting 
at its solvability. In that same work, the exact generating functions 
for punctured staircase polygons with holes of perimeter 
$4$ and $6$ were found. Subsequently in~\cite{guttmann2006j}, the total 
perimeter generating function was found to satisfy 
an 8th order linear ODE by consideration of a large series expansion. The singularity closest to the origin on the positive real axis is at $x=1/4$, coinciding with the location of the staircase polygon and three-choice SAP singularities.  

In~\cite{guttmann2000jwe}, it was noticed that all of the differential 
approximant exponents for the three-choice and punctured staircase 
polygons were equal, and in~\cite{bousquetmelou1999gor} it was seen 
that the inversion relation for their area and anisotropic perimeter 
generating functions were similar. Furthermore, 
in~\cite{guttmann2006j,guttmann2006j2} 
it was noted that the same transfer matrix can be used to enumerate 
the perimeter generating functions of both three-choice and punctured 
staircase polygons, subject simply to different boundary conditions. It 
therefore does not come as a surprise below that we find an exact 
algebraic relationship relating these generating functions.

We note that the exact perimeter generating function for staircase polygons 
with holes in the shape of $ \, 90^{\circ}$-rotated staircase polygons has 
been given and proven in~\cite{jensen2008r} as an algebraic function, the 
solution of a 4th order linear ODE. It appears that there is no relation 
between the rotated-punctured generating function 
and the punctured perimeter generating function considered here.

The punctured staircase polygons total perimeter generating
function series reads
\beq
\label{series2}
P^{\mathrm{P}} \, =\,  \,\, 
 x^8 \,\,  +\, 12 \, x^9  \, +\, 94 \, x^{10}  \, +\, 604 \, x^{11}
   \, +\,3463 \, x^{12}   +\, 18440\,x^{13}\,+\, 93274\,x^{14}
\,\, \, + \,\, \, \cdots  
\eeq

\vskip .2cm 

\section{Differential operator structures}
\label{sec:ODE}

In the following, we denote the order of operators by 
subscripts.  We also denote with $\oplus$ the direct sum 
$\, O_n = \, O_m \oplus O_p$ of 
two order-$m$ and order-$p$ linear differential operators 
$\, O_m$ and $\, O_p$, 
such that all 
solutions of $\, O_m$ and $\, O_p$ are solutions of $\, O_n$. The 
direct sum structure means that the two operators 
$\, O_m$, $\, O_p$ are two possible right-factors of $\, O_n$, namely 
$\, O_n=\, \tilde{O}_m\cdot O_p=\, \tilde{O}_p \cdot O_m$. Conversely, 
forming the operator $\, O_n$ from lower order operators 
$\, O_m$, $\, O_p$ amounts to taking the 
$\, \mathrm{LCLM}(O_m,\, O_p)$ of the two operators, where 
LCLM stands for least common left multiple (see~\cite{put2003s} 
for more details).

The 8th order linear differential operator $\, L_8^{\mathrm{P}}$ 
denotes the operator annihilating 
the perimeter generating function of punctured staircase polygons. 
This  linear differential operator has the following product 
and direct sum decomposition, 
where the product structure is of a different form
  than in~\cite{guttmann2006j}
\beqr
\label{Lmath}
\hspace{-0.95in}&&   \quad  \quad  \quad 
L^{\mathrm{P}}_8 \, \,  =\, \, \, \, 
 I_3\cdot I_1\cdot I_2\cdot \bar{I}_1\cdot \tilde{I}_1 
\, \, \, =\,  \,   \,  J_7\oplus J_1 \,\, =\, \,  \,  
 \left(K_3 \cdot K_2\cdot K_1 \cdot N_1\right)\oplus J_1.
\eeqr

Similarly, for the case of the three-choice staircase 
polygon perimeter generating function 
with imperfect staircase polygons included, we 
have the following 8th order operator product and direct sum 
decomposition
\beqr
\label{Lmath8}
\hspace{-0.95in}&&   \quad  \quad \quad \quad  \quad \quad 
L^{\mathrm{T}}_8 \,\,  = \, \, \,
  L_3\cdot L_2 \cdot L_1\cdot \bar{L}_1 \cdot \tilde{L}_1 
\,\, \, =\, \, \, 
 M_6\oplus M_1 \oplus \bar{M}_1
\nonumber \\ 
\hspace{-0.95in}&&   \quad  \quad \quad 
\quad \quad \quad \quad  \quad \quad 
 \, = \, \,  \, 
\left(N_3\cdot N_2\cdot N_1\right)
  \, \oplus M_1  \,  \oplus  \,  \bar{M}_1, 
\eeqr
And finally, the corresponding 8th order operator for only imperfect 
staircase polygons also decomposes as follows
\beqr
\label{Lmath8b}
\hspace{-0.95in}&&   \quad  \quad \quad \quad \quad \quad  
L^{\mathrm{I}}_8  \, \,  =  \, \,  \,  \,
Q_3 \cdot Q_2 \cdot Q_1 \cdot \bar{Q}_1 \cdot \tilde{Q}_1 
 \, \, \, = \,  \,  \,  M_6 \oplus  \,  R_1 \oplus   \, \bar{R}_1
\nonumber \\ 
\hspace{-0.95in}&&   \quad  \quad \quad 
\quad \quad  \quad \quad  \quad \quad 
  \, = \,  \,\,   \, 
 \left(N_3\cdot N_2\cdot N_1\right)
  \oplus  \,  R_1 \oplus   \, \bar{R}_1. 
\eeqr
Comparing $ \,  L_8^{\mathrm{T}}$ and $ \,  L_8^{\mathrm{I}}$, 
both have direct sum decompositions into the \emph{same} 
6th order linear differential
operator $ \,  M_6= \,  N_3\cdot N_2\cdot N_1$, and simple first order 
operators. As a consequence, these two linear differential 
operators are homomorphic (up to 7th order intertwinners). 

In the product form, operators of the same order for the three 
cases are homomorphic~\cite{put2003s} to each other, for instance 
$\,  I_3 \simeq \,  L_3\simeq Q_3$ 
and $\,  I_2 \simeq \,  L_2\simeq \,  Q_2$. That is, 
$\,  I_3\cdot V_2=\,  W_2\cdot L_3$ for intertwinner operators 
$\,V_2$, $\,  W_2$ of second order.  The 
solutions of the operator $ \, M_6$ which appears both 
in $ \, L_8^{\mathrm{T}}$ and $\, L_8^{\mathrm{I}}$ are 
also solutions of the operator $\, J_7$ 
of $ \, L^{\mathrm{P}}_8$. The $ \,J_7$ and $ \, M_6$ are homomorphic 
to each other with sixth order intertwining operators. The $ \, K_2$ 
and $\,  N_2$ operators are homomorphic to each other with first order 
intertwining operators, and the $\, K_3$ and $ \, N_3$ operators are 
homomorphic to each other with second order intertwining operators 
in one direction, and first order intertwining 
operators in the other direction. Not surprisingly, $L\,_8^{\mathrm{P}}$
is homomorphic to $\,  L_8^{\mathrm{T}}$ and $\,  L_8^{\mathrm{I}}$ 
(again with 7th order intertwinners).

The operator $ \, M_6$ has three solutions analytic at $x= \, 0$, and 
$ \, J_7$ has four analytic solutions. The operators 
$ \, N_3$, $ \, N_2$, $ \, N_1$ have the following form 
($D_x$ denotes, here, and elsewhere in the paper, the differential 
operator $\, \frac{d}{dx}$)
\beqr
\label{N3}
\hspace{-0.95in}&& \,\, \,   
N_3 \, \, =\,\, \,\,   D_x^3  
\, \,\,\,  
+ \frac{p_{24}}{x \cdot \, (1+4x)\,\cdot \,p_{23}} \cdot \, D_x^2
\nonumber\\
\hspace{-0.95in}&&  \quad \quad \qquad  \, \, 
 + \frac{2\,p_{31}}{x^2 \cdot \,(1+4x)\,\cdot \,p_6~q_{24}} \cdot \, D_x
  \, \,\, + \frac{2\,p_{37}}{x^3\cdot\,(1+4x)\cdot\,p_6^2~q_{24}}, 
\\
\label{N2}
\hspace{-0.95in}&&  \,\, \, 
N_2 \,\, =\, \, \, \, \,  D_x^2 \, \,
\, \, \,  - \frac{2\, p_7}{x \cdot \,(1-4x)\,\cdot \,p_6} \cdot \, D_x
\,  \,\, \, \, 
- \frac{2\,p_9}{x \cdot \,(1-4x)^2\,( 1+x+7x^2)\,\cdot \,p_6},
 \\
\hspace{-0.95in}&&  \,\, \, 
\label{N1}
N_1 \,\, =\, \,\, \,\,   D_x  \,\,\,\,  + \frac{4}{1\, -4x}.
\eeqr
where the polynomials $ \, p_j,\,q_j$ of order $ \, j$ 
are given in ~\ref{app:operators}. The 
solution of $ \, N_1$ is $ \, (1-4x)$.

Since the $\, L_8^{\mathrm{T}}$ and $ \, L_8^{\mathrm{I}}$ 
operators are quite similar, as noted previously, it is 
not surprising that their LCLM only has 
order 10 instead of the generic 16th order expected of the LCLM 
of two 8th order operators. Similarly, 
that is the case among any pair of the operators 
$ \, L_8^{\mathrm{T}}$, $ \, L_8^{\mathrm{I}}$, $ \, L_8^{\mathrm{P}}$. To some 
extent this explains that the LCLM of all three operators, 
which encapsulates the generating functions of three-choice, imperfect, 
and punctured staircase polygons, is only of 12th order, half of the 
expected order. See \ref{app:LCLM} for details of the LCLM 
structures. Note that any linear combination of the form
\beqr
\label{order1sols}
\hspace{-0.95in}&&   \quad  \quad \quad \quad \quad \, \,
A_0 \,\,\, + \frac{A_1}{(1\,-4x)} \,\, \,
+ A_2 \cdot \, (1\,-4x) \,\,\, \,+ A_3\cdot \, (1\,-4x)^2 \, 
\nonumber \\
\hspace{-0.95in}&&   \quad  \quad \quad \quad \quad \quad \quad \, \, \,
+ \frac{A_4}{\sqrt{1\,-4x}}  \,  \,   \, \,
+ A_5 \cdot \,\sqrt{1\,-4x} \,  \,\, \,\, + A_6\cdot \,(1\,-4x)^{3/2}, 
\eeqr
is actually solution of the 12th order LCLM of the three 8th order
linear differential operators $ \, L_8^{\mathrm{T}}$, $ \, L_8^{\mathrm{I}}$, 
and $ \, L_8^{\mathrm{P}}$.

\subsection{Equivalence of generating functions}
\label{equivgener}

From~\cite{conway1997gd}, the relationship between the three-choice 
and imperfect staircase polygons is known to be
\beqr
\label{imp_to_three}
\hspace{-0.95in}&&   \quad  \quad \quad  \quad \quad 
\frac{1}{2}\,P^{\mathrm{T}} 
\, \, -  P^{\mathrm{I}} \, \, = \,  \, \, \,
x \cdot \, \frac{dP^{\mathrm{S}}}{dx} 
\, \,  \, = \, \, \, \left(\frac{x}{\sqrt{1-4x}}\, \, -x \right), 
\eeqr
where $\, P^{\mathrm{S}}$ is the staircase polygon generating 
function in~(\ref{Px}).

The right-hand side of~(\ref{imp_to_three}) is a solution of a 2nd order
operator which is the LCLM of two simple first order operators. From 
equation~(\ref{imp_to_three}) one immediately deduces that $\, P^{\mathrm{T}}$ 
is a solution of the LCLM of $\, L_8^{\mathrm{I}}$ and of this 2nd order operator, 
yielding the result that the LCLM of $\, P^{\mathrm{T}}$ and $\, P^{\mathrm{I}}$ is 
of 10th order as seen in \ref{app:LCLM}. Conversely, the calculations 
of \ref{app:LCLM} provide a means 
to deduce the relationship of equation~(\ref{imp_to_three}).

Similarly, from the LCLM of $ \,P^{\mathrm{I}}$ 
and $ \,P^{\mathrm{P}}$ in \ref{app:LCLM}, 
one can find the following algebraic relationship 
\beqr
\label{imp_to_punc}
\hspace{-0.95in}   \quad   \quad   \quad   \quad   \quad  
P^{\mathrm{I}} \, \, +P^{\mathrm{P}} \, \, 
\,\,    &=&  \, \, \, 
  - \frac{x^2}{2} \cdot \,\frac{dP^{\mathrm{S}}}{dx}
\,  \,+  \,\,  \frac{x^3}{1\, -4x}
 \\
\hspace{-0.95in}   \quad  \quad  \quad  
 \quad   \quad   \quad     \quad    \quad     \quad  
&=&~~  \, \frac{x}{2}\cdot \left(x-\frac{x}{\sqrt{1-4x}} \right)
\, \,  \,+  \,\,  \frac{x^3}{1\, -4x}.
\eeqr
Using the two relationships~(\ref{imp_to_three}) and (\ref{imp_to_punc}), 
we can deduce the following relationship
\beqr
\hspace{-0.95in}&&   \quad   \quad   \quad   \quad    
P^{\mathrm{P}}\,  +\frac{1}{2}\,P^{\mathrm{T}} 
\,\, =\, \,\,  \, 
-\frac{x\cdot\, (x\, -2)}{2} \cdot \, \frac{dP^{\mathrm{S}}}{dx} 
 \,\, \,  \, + \,\,  \frac{x^3}{1\, -4x}
\\
\hspace{-0.95in}&&   \quad  \quad  \quad   \quad  
\qquad\qquad~\, =  \,  \,\,  \, 
\frac{x \cdot \, (x\, -2)}{2} \cdot \left(x\, -\frac{x}{\sqrt{1-4x}} \right) 
~\,+ \,\,  \frac{x^3}{1\, -4x}.
\label{punc_three}
\eeqr
While the LCLMs of the operators $\, L_8^{\mathrm{T}}$, 
$\, L\, _8^{\mathrm{I}}$, $\, L_8^{\mathrm{P}}$ provide a proof 
of the above relationships, we still have not found a direct combinatorial
 derivation or interpretation of relationships 
(\ref{imp_to_punc})--(\ref{punc_three}).

Note that by eliminating $ \, dP^\mathrm{S}/dx$ 
between (\ref{imp_to_three})--(\ref{punc_three})
 and thereby eliminating all square roots, we find the following 
very simple relationship among $\, P^{\mathrm{T}}$, $\, P^{\mathrm{I}}$, 
$\, P^{\mathrm{P}}$
\beqr
\hspace{-0.95in}&&   \quad  \quad  \quad \quad  \quad 
P^{\mathrm{P}}\, \,  + \left(\frac{x}{4}\right) \cdot\, P^{\mathrm{T}} 
\, \, + \left(1-\frac{x}{2}\right)\cdot\, P^{\mathrm{I}}
 \,\, \, \,  = \, \, \, \,  \frac{x^3}{1\, -4x}.
 \label{imp_punc_three} 
\eeqr
Again, we have not yet found a combinatorial explanation 
of (\ref{imp_punc_three}).


\section{Results}
\label{sec:results}

The solutions to all three generating functions can be most simply expressed 
in terms of the solutions in equation (\ref{order1sols}), which are 
the seven first order solutions of 
$ \, L_{12}^{\mathrm{TIP}}$
$=\, \mathrm{LCLM}(L_8^{\mathrm{T}},L_8^{\mathrm{I}},L_8^{\mathrm{P}})$,
 plus solutions of
$ \, M_6= \, N_3\cdot N_2\cdot N_1$. We here focus on the solutions of the 
linear differential operator $ \, L_8^{\mathrm{I}}$, since the solutions 
of the other two generating functions are easily related to the solution 
of $\,L_8^{\mathrm{I}}$ by relationships~(\ref{imp_to_three})
and (\ref{imp_to_punc}). 

The generating function for imperfect staircase polygons is given 
as the sum of algebraic and transcendental functions
\beqr
\label{algtrans}
\hspace{-0.95in}&&   \quad  \quad  \quad 
\quad \quad \quad \quad \,\,
P^{\mathrm{I}} \,  \,  = \, \,  \, 
\frac{1}{60}\cdot\left(P^{\mathrm{I}}_{\mathrm{alg}}\, 
 + P^{\mathrm{I}}_{\mathrm{trans}}\right), 
\eeqr
where the algebraic part $ \, P^{\mathrm{I}}_{\mathrm{alg}}$ is actually 
a series with \emph{integer} coefficients 
\beqr
\hspace*{-\mathindent}&&\quad
P^{\mathrm{I}}_{\mathrm{alg}} \,\,  = \,\,\,\,   \,   
\frac{135}{8}\, \,\,  \,   + \frac{59}{4}\cdot \,(1\, -4x)
 \,\,  \, \,   + \frac{15}{8} \cdot \,(1\, -4x)^2 
\,\,\,  \,   -\frac{85}{16}\cdot\frac{1}{\sqrt{1\, -4x}} 
\nonumber\\
\hspace*{-\mathindent}&&   \quad \quad 
\quad \quad \quad \quad \,\,
-\frac{105}{8} \cdot \,\sqrt{1\, -4x}
 \, \,\, \,\,   \,
 -\frac{65}{16} \cdot \,  (1\, -4x)^{3/2}
 \\
\hspace*{-\mathindent} && \quad \,\,
= \,  11\, -34 x\, -70 x^3 -265x^4 -1020x^5 -3920x^6 
 -15060x^7-57915x^8 \, +    \,  \ldots
\eeqr
and where the transcendental part $\, P^{\mathrm{I}}_{\mathrm{trans}}$
\beq
\label{PIser}
P^{\mathrm{I}}_{\mathrm{trans}} = \, 
-11+\, 34x +70x^3 +325x^4 +1380x^5 +5660x^6
 +22860x^7 +91575x^8 \, + \ldots
\eeq
is a solution of $ \,N_3 \cdot \,N_2 \cdot  \,N_1$ and 
can be decomposed as the linear combination 
\beqr
\hspace{-0.95in}&&   \quad  \quad  \quad \quad \quad \quad \quad
P^{\mathrm{I}}_{\mathrm{trans}} \,\,  =\, \, \, \, 
 -\frac{19}{2}\cdot \, \mathrm{Sol}_2 \,\,\,   
- \frac{3}{2}\cdot\,  \mathrm{Sol}_3, 
\eeqr
of the two regular solutions $ \,\mathrm{Sol}_2$ 
and $ \,\mathrm{Sol}_3$
\beqr
N_2 \cdot  \,N_1 \left(\mathrm{Sol}_2 \right)  \,= \, \,  0, 
\label{Sol2def}  \\ 
N_3\cdot \, N_2 \cdot \, N_1 \left(\mathrm{Sol}_3 \right)  \,= \, \, 0,
 \label{Sol3def}
\eeqr
corresponding to the series expansions with \emph{integer} coefficients
\beqr
\label{Sol2}
\hspace*{-\mathindent}\quad \mathrm{Sol}_2 &=& \, \, 
\,  1\,\,  -2x\, +3x^2 +4x^3 +13x^4 +36x^5 
+95x^6 +246x^7 +588x^8 \,\,  +  \,\ldots 
\\
\label{Sol3}
\hspace*{-\mathindent}\quad \mathrm{Sol}_3  &=& \, \, 
\, 1\,\,  -10x -19x^2 -72x^3 -299x^4 -1148x^5
 -4375x^6 -16798x^7 \, \, +  \,\ldots 
\eeqr

The series $\, \mathrm{Sol}_2$, $ \, \mathrm{Sol}_3$ are the 
unique series, up to an overall factor, multiplying the 
largest logarithmic power of the formal solutions of 
 $\, N_2 \cdot  \,N_1$ and $\, N_3\cdot\,N_2 \cdot  \,N_1$, 
respectively, as seen in equation (\ref{SSS}) 
of \ref{app:series}.  

It is remarkable to observe that the linear combination of rational 
coefficients from $(1/60)\, P^{\mathrm{I}}_{\mathrm{alg}}$,
 $(19/120)\, \mathrm{Sol}_2$ and $(3/120)\, \mathrm{Sol}_3$ 
actually  gives the integer series corresponding to 
$\, P^{\mathrm{I}}$. As will be seen below, $\, N_2$ and $\, N_3$ 
are of a quite different nature, so that it is rather surprising 
that a solution of $ N_3 \cdot  N_2  \cdot N_1$
is precisely able 
to compensate the $\, P^{\mathrm{I}}_{\mathrm{alg}}/60$ series 
in order to generate the integer series of $\, P^{\mathrm{I}}$.

\subsection{Exact $\, N_2$ solution}
\label{exactN2}

The second order linear differential operator $ \, N_2$ has the following 
solution as a $\, _2F_1$ hypergeometric function with a rational 
cubic pullback, which can be found, for example, using the program 
hypergeomdeg3 described in~\cite{kunwar2013h} 
\beqr
\label{N2sol}
\hspace{-0.95in}&&   
\quad 
\mathrm{Sol}(N_2) \,\,  = \,\,\,   {{ 1\,+x\,+7 \,x^2 } \over {
18 \cdot \, x \cdot \, (1-x)^2 \cdot \, (1 \, -4 \,x)^{3/2} }} 
\cdot \, {\cal S} 
\\
\label{sersolN2}
\hspace{-0.95in}&&  \quad  \quad 
\, \, = \, \, \, \,  \,
1 \, \, \, +7\,x \, \, \, +28\,{x}^{2} \, +122\,{x}^{3} \, +500\,{x}^{4}
 \, \,+1997\,{x}^{5}\, \, +7899 \,{x}^{6}\,
+30996\,{x}^{7}
\nonumber\\
\hspace{-0.95in}&&  \quad \quad \quad \quad \quad 
 \,\,\, +120774\,{x}^{8}\, \, +468035\,{x}^{9}\, \, 
+1805351\,{x}^{10} \,\, +6932732\,{x}^{11} \,\, 
 \, +  \,  \, \cdots
\eeqr
where 
\beqr
\label{N2solS}
\hspace{-0.95in}&&  
{\cal S} \, \,= \, \, \,
 (1-x)\, (1 \, -4 \,x) \, (1 \, + 45\, x^2 \, + 44 \, x^3) \cdot \, 
{{d {\cal H}} \over {dx}}\,
 \, \, + \, \, 18 \cdot \, (1 \, -3 \, x \, -13 \, x^2) \cdot \, {\cal H}, 
\eeqr
with
\beqr
\label{N2solH}
\hspace{-0.95in}&&   \quad \quad \quad \quad \quad 
\quad 
{\cal H} \,\, \,  = \, \, \, \, 
 _2F_1\left(\left[{{1} \over {3}}, \, {{2} \over {3}}\right], 
\, [1], \, \frac{27x^3}{(1-x)^3}\right). 
\eeqr
This solution can also be expressed as the following
 sum of two contiguous $\, _2F_1$ hypergeometric functions
\beqr
\label{N2solcontiguous}
\hspace{-0.95in}&&  \, 
\mathrm{Sol}(N_2) \,  = \,\,
\frac{1}{(1-x)^2 \cdot \, 
 (1-4x)^{3/2}} \cdot \,\left[(x+45x^3+44x^4) \cdot \, 
~{}_2F_1\left(\left[\frac{1}{3},\frac{2}{3} \right],[2],
 \, \frac{27 x^3}{(1-x)^3}\right)\right.
 \nonumber\\
\hspace{-0.95in}&&\quad \quad  \quad \, \,\,
\left. +~(1\,+x\,+7x^2) \cdot \, (1\,-3x\,-13x^2) \cdot \, 
~{}_2F_1\left(\left[\frac{1}{3},\frac{2}{3} \right],[1],
 \, \frac{27x^3}{(1-x)^3}\right)   \right].
\eeqr

As a sum of two contiguous $\, _2F_1$ hypergeometric functions, 
we can wonder whether a different contiguous basis exists 
for $\, \mathrm{Sol}(N_2)$ which gives smaller algebraic 
pre-factors. We have made use of the Maple 
procedure contiguous2f1.mpl developed by 
Vid{\=u}nas~\cite{vidunas2002}, but 
we have been unable to find a simpler contiguous basis. 

While the form of the pullback in (\ref{N2solH}) does not 
reveal the physical singularity at $\, x=\, 1/4$, the standard 
hypergeometric identity
\beqr
\label{hypergident}
\hspace{-0.95in}&&  \quad \quad  \quad \quad \,\,
{}_2F_1([a,b],[c],z)\, )\, \,  = \,\, \, \, 
(1-z)^{-a} \cdot \,  {}_2F_1\left([a,c-b],[c],\frac{z}{z-1} \right), 
\eeqr 
changes the pullback to
\beqr 
\label{pullchange}
\hspace{-0.95in}&&  \quad \quad  \quad \quad \,\,
z \,=\, \, {\frac {27{x}^{3}}{ (1\, -x)^{3}}}   
\, \,  \, \, \longrightarrow \,\,  \, \,\, \,\, \, 
 \frac{z}{z-1}\,\, = \,\, \, -\frac{27x^3}{(1-4x)(1\, +x\, +7x^2)}, 
\eeqr 
from which we see that the physical singularity is mapped to 
$ \, z=\, \infty$ and where we also see the appearance of the 
unphysical singularities at the roots of polynomial
$\, 1\, +x\, +7x^2$, which were already observed 
in~\cite{guttmann2006j}.

\subsection{Exact $\, N_3$ solution}
\label{exactN3}

The solution to the operator $\, N_3$ is much more involved and the 
path to discovering its solution is not obvious. The discovery of 
the solution starts with seeing that the exterior square 
of $ \, N_3$ has a rational solution, 
which means that this 3rd order operator is homomorphic to the
{\em symmetric square} of a second order operator. Constructing 
$ \, \bar{N}_3$ by conjugating $ \, N_3$ by the square root of 
this rational function, we 
used the ``conic program" described in~\cite{hoeij2006c} and available at~\cite{Conic} in order to find a 
second order linear differential operator $ \, \bar{V}_2$, such that 
its {\em symmetric square} is, not equal, but
 {\em  homomorphic} to $ \, \bar{N}_3$. This second order 
linear differential operator $ \, \bar{V}_2$ reads 
\beqr
\hspace{-0.95in}&&  \quad \quad \quad  \,  \, \, \, 
\bar{V}_2 \,\,   =  \, \,   \, \, 
D_x^2 \,  \,  \, \, 
+{\frac { p_{36} }{ 
(1 \, - 4\,x)  \,
 (1 \, + 4\,x)  \, (1\, +4\,{x}^{2}) \cdot \, p_{33} }}
 \cdot \, D_x 
\nonumber \\ 
\hspace{-0.95in}&&  \qquad  \qquad \quad 
\, \,  \, \,  \,\,   \, \,   \,  \,  \,  \,  \,  
- 2 \cdot \, 
{\frac { q_{36}}{
x \cdot \,  (1 \, - 4\,x)  \, (1 \, + 4\,x) 
 \, (1\, +4\,{x}^{2}) \cdot \, p_{33} }},  
\eeqr
where the polynomials $\, p_{33}$, $\, p_{36}$ and $\, q_{36}$  
 are given in \ref{app:operators}.
The quite large degree $\, 33$ polynomial $\, p_{33}$ corresponds 
to { \em apparent singularities}. To go further it is crucial to 
get rid of $\, p_{33}$ and to find a 2nd order linear differential operator homomorphic 
to $ \, \bar{V}_2$ with none of these apparent singularities. This 
corresponds to the so-called 
{\em desingularization} of a Fuchsian linear differential 
operator. Note that in general it is {\em not easy} to get rid of the apparent 
singularities {\em without introducing new ones}. We are
not interested in such a ``partial'' desingularization, but in a 
``complete'' desingularization, which is, in general,
 {\em not always possible} without increasing the order. 
One looks for an operator equivalence that keeps the exponents of the 
new linear differential operator at the true singularities 
in a fairly narrow range: to ``minimize'' the number of apparent 
singularities one needs to ``maximize'' the sum of all the exponents 
at all the true regular singularities\footnote[1]{The technical
details on how to implement these briefly sketched ideas will be
found in a forthcoming paper~\cite{ErdalHoeij}.}.
In this particular case, we have been able to find such a
2nd order Fuchsian linear differential operator $\, V_2$, 
homomorphic to $ \, \bar{V}_2$, with no apparent singularities\footnote[2]{This
much simpler second order operator can be obtained from van Hoeij's program ReduceOrder available here~\cite{ReduceOrder}.} .  
This  much simpler
second order operator $\, V_2$ reads
\beqr
\hspace{-0.95in}&&  \quad  \quad 
(1\, -4\,x)^{2} \cdot \, (1\, +4\,x)^{2}
 \cdot \, (1\, +4\,{x}^{2})^{2} \cdot \, {x}^{2} \cdot \, V_2
\nonumber \\ 
\hspace{-0.95in}&&  \quad \quad \quad  \quad 
 \, =\, \,  \, 
(1\, -4\,x)^{2} \cdot \, (1\, +4\,x)^{2}
 \cdot \, (1\, +4\,{x}^{2})^{2} \cdot \, {x}^{2} \cdot \, D_x^{2}
\nonumber \\ 
\hspace{-0.95in}&&  \quad \quad \quad  \quad 
  \quad \, 
\,\, + \, (192\,{x}^{4}+24\,{x}^{2}-1) \cdot \, 
(1 -4\,x)  \cdot \, (1\, +4\,x) 
 \cdot \, (1\, +4\,{x}^{2}) \cdot \, x  \cdot D_x
\nonumber \\
 \label{U2}
\hspace{-0.95in}&&  \quad \quad \quad  \quad 
  \quad
\, \,\, +16128\,{x}^{8} +3280\,{x}^{6} +532\,{x}^{4} -16\,{x}^{2} +1.
\eeqr
These two operators, $\, \bar{V}_2$ and $\, V_2$,
are homomorphic with first order 
intertwinners
\beqr
\label{V2A1B1V2bar}
\hspace{-0.95in}&&  \quad  \quad \quad \quad \, \,
 V_2 \cdot \, A_1 \, =\, \, B_1 \cdot \, \bar{V}_2, 
\qquad \quad   
C_1  \cdot \, V_2  \, =\, \, 
\bar{V}_2 \cdot \, D_1,  
\eeqr
where\footnote[1]{Note that the 3rd order operators 
$\, V_2 \cdot \, A_1$ and $\, C_1  \cdot \, V_2$ 
in (\ref{V2A1B1V2bar}) still have $\, p_{33}$
(and another polynomial $\, \tilde{p}_{15}$, see below) 
as apparent singularities.} the first order intertwiners $\, A_1$, $\, B_1$, $\, C_1$, $\, D_1$
are of the form
\beqr
\hspace{-0.95in}&&  \qquad 
A_1 \, \, =\, \, \,\, 
 {{x \cdot \,  (1  \, -4\,x)\, (1\, +4\,x)\, (1\, +4\,x^2)
} \over {p_{33}}} 
\cdot \, 
\Bigl( \tilde{p}_{15} \cdot \, D_x \, \,
 -2\cdot \,  \tilde{q}_{14} \Bigr),  
\label{A1}
\\ 
\hspace{-0.95in}&&  \qquad
B_1 \,\,  =\, \,\, \, {{1} \over {p_{33}}}
\cdot \, \left(  \tilde{p}_{15}
 \cdot \, (1  \, -4\,x) 
\, (1\, +4\,x)\, (1\, +4\,x^2) \cdot \, x \cdot \, D_x
 \, \, +{{ \tilde{p}_{52}} \over { p_{33} }} \right), 
\label{B1}
\\
\label{C1}
\hspace{-0.95in}&&  \qquad
C_1 \,\,  =\, \,\, \,
 \tilde{p}_{15} \cdot \, \left(D_x \, \,- {{d \, \ln\Bigl(1/U
\Bigr)} \over {dx}} \right)
\, \, + \, \, {{ \tilde{p}_{47}} \over { p_{33} }}, 
\\
\label{D1}
\hspace{-0.95in}&&  \qquad
D_1 \, \, =\, \, \, \,
 \tilde{p}_{15} \cdot \,
 \left(D_x \, \,- {{d \, \ln( x\cdot U)} \over {dx}} \right)
\,\, + \, \, 2 \cdot \, \tilde{p}_{14}. 
\eeqr
where $\, U$ denotes the algebraic function 
$\, \sqrt {(1 \, -16\,x^2) \, (1 \, +4\,x^2)}$, 
and where the polynomials 
 $ \tilde{p}_{14}$, $ \tilde{q}_{14}$, $ \tilde{p}_{15}$, $ \tilde{p}_{47}$
and $ \tilde{p}_{52}$ are given in \ref{app:operators}. Note 
that $\, \tilde{p}_{14}$ or $\, \tilde{q}_{14}$ 
in (\ref{A1}), (\ref{D1}) do not correspond to derivatives 
of $\, \tilde{p}_{15}$. Therefore the first order intertwiners $\, A_1$ 
and $\, D_1$ {\em are not Fuchsian operators}, since one does not have a 
logarithmic derivative. Along this line it is obvious that $\, B_1$ 
and $\, C_1$ are also {\em not Fuchsian operators}: these four 
order-one intertwiners, corresponding to the homomorphisms the two
{\em Fuchsian} order-two operators, $ \, \bar{V}_2$ and $ \,V_2$ , 
are {\em not themselves Fuchsian}. 

The 2nd order operator  $\, V_2$ being homorphic to $ \, \bar{V}_2$,
and the 3rd order operator $\, N_3$ being homorphic to 
the symmetric square of $ \, \bar{V}_2$, one finds straightforwardly 
that  the 3rd order operator $\, N_3$ is homomorphic to
the symmetric square of the (much simpler) 2nd order 
operator $\, V_2$
\beqr
\hspace{-0.95in}&&  \qquad \qquad  \qquad \quad 
N_3 \cdot \, T_2  
\,\, =  \, \, \, W_2 \cdot  \,\mathrm{Sym}(V_2)^2, 
\eeqr
where the intertwinner $\, T_2$ is a 2nd order operator of the form
\beqr
\hspace*{-\mathindent}&& \quad \, \, T_2 \, = \, \, \, 
\frac{4\,p_{10}}{3\, x^3 
\cdot \,(1\,-4x)^2 \,(1\,+x\,+7x^2) \cdot \,p_6}\cdot \, D_x^2
\nonumber\\
\hspace*{-\mathindent}&&\quad \quad \quad \, \, 
+ \frac{4\,p_{14}}{3\, x^4 
\cdot \, (1\,-4x)^2 \,(1\,+x\,+7x^2)\,(1-12x^2-64x^4) \cdot \,p_6}
\cdot \, D_x
\\
\hspace*{-\mathindent}&&\quad  \quad \quad \, \, 
+\frac{16\,p_{18}}{3\, x^5\cdot \, 
(1 \,-4x)^2 \,(1\, +x\, +7x^2) \,
 (1-24x^2 +16x^4 +1536x^6 +4096x^8) \cdot \,p_6},
\nonumber
\eeqr
with polynomial coefficients $\, p_j(x)$ of order $\, j$ defined 
in~\ref{app:operators}. 

The relevant solution of $\, V_2$ is given by 
\begin{eqnarray}
\hspace{-0.95in}&& 
\mathrm{Sol}(V_2)  
\,  \,= \, \, \,\,  {\cal S}(x, \, U) \, \,  \,= \, \,
 \nonumber \\ 
\label{solN3} 
\hspace{-0.95in}&&  
\, \, x \cdot \, U \cdot \, 
\left({{13 \,-28\,{x}^{2}\,  -12\,U} \over {
 (1 + \, 20\,{x}^{2})^2 }}\right)^{1/4} \cdot \,
 _2F_1\left(\left[{{1} \over {8}}, {{3} \over {8}}\right], \, [1], \, 
\, {\frac {4096\,{x}^{10}}{ (1 -4\,{x}^{2} \, +U)^{4}}}\right)
 \\ 
\label{HeunGsol}
\hspace{-0.95in}&&  
\,  \,= \, \, \,\,  x \cdot \, U
 \cdot \,  
\mathrm{Heun}\Bigl( -{{1} \over {4}}, \, {{1} \over {16}}, \, \,
 {{3} \over {8}}, \, {{5} \over {8}},\, 1,
 \, {{1} \over {2}}, \,  -4 \, x^2\Bigr) 
  \\ 
\label{sersolV2}
\hspace{-0.95in}&&  \, \,  \,  
 \quad  \, \, \,=   \,  \, \,\,\, x \,\,  \,  \,
  -5\,{x}^{3} \, \,  -{\frac {95}{2}}\,{x}^{5} \,\,  \, 
-{\frac {655}{2}}\,{x}^{7} \, \,
  -{\frac {27365}{8}}\,{x}^{9} \, \, \, 
-{\frac {305131\,{x}^{11}}{8}} \, \,
-\frac {7365195}{16} \,{x}^{13}
 \nonumber \\ 
\hspace{-0.95in}&& \quad  \quad  \quad  \quad \,  \,   \, \,  \, 
-{\frac {92787415}{16}} \, {x}^{15} \, \,  \, 
-{\frac {9671421805}{128}}\,{x}^{17} \, \, \, 
-{\frac {129164164935}{128}}\, {x}^{19} \, 
\, \,\,\,  +  \, \, \,  \cdots  
\end{eqnarray}
with
\beqr
\label{Udef}
\hspace{-0.95in} \quad\quad \quad  \quad \quad  \, \,  \quad  \, 
U \, \, \, = \, \, \, \
+ \, \sqrt{(1\,  -16 \,x^2)  \, (1\, +4\,{x}^{2})}.
\eeqr

Finding the solution of $\,V_2$ in (\ref{solN3}) as a $\, _2F_1$ 
hypergeometric function with an {\em algebraic} pullback is highly 
non-trivial. It can be found, for example, using the new Maple procedure 
hypergeometricsols written by Erdal Imamoglu 
and available at~\cite{imamoglu2015}. The result found from the program 
contains a different algebraic pre-factor as well as a different 
algebraic pullback, which can then 
be simplified to the form (\ref{solN3}) found above\footnote{Similarly, 
the solution of $\, N_2$ in (\ref{N2sol}) can be found in an alternative 
form through the hypergeometricsols program.}.

One can of course imagine that considering the other 
branch of the square root in (\ref{Udef}) gives an alternate expression. 
This is actually
the case, changing  $\, U$ 
into $\, -U$ gives, in fact, the same solution
{\em up to a $\, -5^{-1/2}$ factor}, in the following alternative form
\begin{eqnarray}
\hspace{-0.95in}&& 
\label{othersolN3} 
\mathrm{Sol}(V_2)  \,  \,= \, \, \,\,\, 
  -5^{-1/2} \cdot \, {\cal S}(x, \, -U) \, \, \, \,= \, \,
 \nonumber \\ 
\hspace{-0.95in}&&  \quad 
  x \cdot \, U \cdot \, 
\left({{13 \,-28\,{x}^{2}\,  +12\,U} \over {
 5^2 \cdot \, (1 + \, 20\,{x}^{2})^2 }}\right)^{1/4} \cdot \,
 _2F_1\left(\left[{{1} \over {8}}, {{3} \over {8}}\right], \, [1], \, 
\, {\frac {4096\,{x}^{10}}{ (1 -4\,{x}^{2} \, -U)^{4}}}\right).
\end{eqnarray}

Note that the two pullbacked hypergeometric functions in  
(\ref{solN3}) and  (\ref{othersolN3}) 
are actually two different series with {\em integer coefficients}
\begin{eqnarray}
\label{expand1}
\hspace{-0.95in}&& \quad  
 _2F_1\left(\left[{{1} \over {8}}, {{3} \over {8}}\right], \, [1], \, 
\, {\frac {4096\,{x}^{10}}{ (1 -4\,{x}^{2} \, +U)^{4}}}\right)
 \, \, \, = \, \, \, \,\,
1 \, \,\,+12\,{x}^{10} \,\,+240\,{x}^{12}\, \,+4200\,{x}^{14}
 \nonumber \\
 \hspace{-0.95in}&&  
\quad \quad \quad \, \,\,\, \,+67200\,{x}^{16} \,
+1040700\,{x}^{18} \,+15830388\,{x}^{20} \,+238737720\,{x}^{22} 
\,\,\,+\,\, \cdots 
\end{eqnarray}
while
\begin{eqnarray}
\label{expand1}
\hspace{-0.95in}&&  \quad  
 _2F_1\left(\left[{{1} \over {8}}, {{3} \over {8}}\right], \, [1], \, 
\, {\frac {4096\,{x}^{10}}{ (1 -4\,{x}^{2} \, -U)^{4}}}\right)
 \, \, \, = \, \, \, \,\,
1\,\,\,+12\,{x}^{2}\,-12\,{x}^{4}\,+744\,{x}^{6}\,
\nonumber \\ 
\hspace{-0.95in}&&  \quad \quad 
\quad\,\,
 -2700\,{x}^{8} \,+115140\,{x}^{10}\,-782520\,{x}^{12}\,
+24418920\,{x}^{14}\,-238316940\,{x}^{16}
\nonumber \\ 
\hspace{-0.95in}&& \quad \quad \quad  \,\,
\,+6113609700\,{x}^{18}\, -74768429700\,{x}^{20}\,
+1698621342600\,{x}^{22} \,\, \, \,+\,\, \, \cdots 
\end{eqnarray}
The series solution (\ref{sersolV2}) is not a series with 
integer coefficients but it is globally bounded~\cite{bostan2013bchm}
so that it can be recast into a series with integer coefficients by
changing $\, x\, \rightarrow \, 2 \, x$. In fact, the 
square of the series solution (\ref{sersolV2}) is  a series 
with integer coefficients, since the square of the algebraic 
pre-factor is a series with integer coefficients.

\vskip .1cm

The solution of $ \, N_3$ is straightforwardly found by applying the 
second order  intertwinner operator $ \, T_2$ to the square 
of the solution of $\, V_2$
\beqr 
\label{N3toU2hom}
\hspace{-0.95in}&&  \qquad \qquad \qquad 
\mathrm{Sol}(N_3)\,\,  = \, \,\, T_2\big(\mathrm{Sol}(V_2)^2 \big).
\eeqr
Using the {\em Clausen identity}~\cite{Hardy}
\beqr 
\label{Clausen}
\hspace{-0.95in}&&  \quad \quad \qquad 
 {}_2F_1\left(\left[{{1} \over {8}}, \,  {{3} \over {8}}\right], 
\, [1], \, z \right)^2
 \,\,  = \, \,\, 
_3F_2\left(\left[{{1} \over {4}}, \, {{1} \over {2}},
 \,  {{3} \over {4}}\right], \, [1, \, 1], \, z \right), 
\eeqr
one can rewrite the quadratic expression (\ref{N3toU2hom}) 
of the $\, _2F_1$ in (\ref{solN3}) in terms of 
the pullbacked $\, _3F_2$ and its derivatives
\beqr 
\label{pullbacked3F2}
\hspace{-0.95in}&&  \qquad \quad \qquad 
_3F_2\left(\left[{{1} \over {4}}, \, {{1} \over {2}}, \, 
 {{3} \over {4}}\right], \, [1, \, 1], \, \, 
{\frac {4096\,{x}^{10}}{ (1 -4\,{x}^{2} \, +U)^{4}}} \right). 
\eeqr

\vskip .1cm

We further note that using the following 
hypergeometric identity
\beqr
\hspace{-0.95in}&& \quad 
{}_2F_1\left(\left[\frac{1}{8},\frac{3}{8}\right],[1],\, 
\frac{16 \, z^2 \cdot \, (1-z)}{(2-z)^4}\right) 
\, \, = \, \, \, 
 \left(\frac{2-z}{2}\right)^{1/2} \cdot 
 \, {}_2F_1\left(\left[\frac{1}{2},\frac{1}{2}\right],[1],z \right), 
\eeqr
it is possible to transform the $\, _2F_1$ hypergeometric 
function in $\, \mathrm{Sol}(N_3)$ to the 
complete elliptic integrals of the first and second kinds 
$\, K(z)$, $\, E(z)$.

In \ref{integrating} we consider the problem of integrating 
$ \, \mathrm{Sol}(N_3)$ back through $ \, N_2\cdot N_1$ 
in order to find $ \, \mathrm{Sol}_3$ of (\ref{Sol3def}).

\subsection{Singularity Analysis}
\label{singanaly}
The nearest singularity on the positive real axis for the generating 
functions is at $x=1/4$. It was already known from~\cite{guttmann2006j} 
and~\cite{guttmann2006j2}) that the singularity at $x=1/4$ has a square root 
divergence as well as a logarithmic singularity. The square root divergence
has contributions from both the algebraic and transcendental parts of the 
generating function solution in (\ref{algtrans}). From the Heun function 
form of $ \, \mathrm{Sol}_3$ in (\ref{HeunGsol}) we can see clearly 
that $ \, \mathrm{Sol}_3$ only contributes corrections to the square 
root singularity of the algebraic part. Therefore, the sole contribution 
to the logarithmic singularity at $x=1/4$ comes from the hypergeometric 
solution $\mathrm{Sol}_2$ in (\ref{N2sol}) with (\ref{pullchange}), 
which also contributes to the square root singularity.

Note that the singularities at $\, 1\,+4x \,= \, 0$, 
$\, 1\,+4x^2\,= \, 0$, $ \, 1+x+7x^2\,= \, 0$, analyzed 
in~\cite{guttmann2006j} 
and~\cite{guttmann2006j2}) 
only emerge from the transcendental part 
of the solution in (\ref{algtrans}).


\section{Modular forms and hypergeometric identities}
\label{sec:identities}

\subsection{The solutions of $\, N_2$ as modular forms}
\label{sec:identitiesN2}

The solution (\ref{N2sol}) of $ \, N_2$ in terms of (\ref{N2solH}) 
can actually be seen to be associated with a {\em modular form}. 

Let us consider the {\em modular curve}
\beqr
\label{modular}
\hspace{-0.95in}&& 
\,10077696 \cdot \,{C}^{3}{D}^{3} \,
 +3779136 \cdot \,{C}^{2}{D}^{2} \cdot \, (C+D)
 \,\, \,  +472392 \cdot \,C\, D \cdot \, ({C}^{2} -87\,CD +{D}^{2})
\nonumber \\ 
\hspace{-0.95in}&& \quad \quad \quad 
 \,\, +19683 \cdot \, (C+D) 
 \cdot \, ({C}^{2}+440\,CD+{D}^{2}) \, \, 
 -59049 \cdot \,({C}^{2} -87 \,CD \, +\,{D}^{2}) 
\nonumber \\ 
\hspace{-0.95in}&& \quad \quad \quad  \quad  \quad 
\, \,\, \, +59049 \cdot \,(C+\,D)\, \, \,  -19683 \, 
\, \, \,  = \, \,  \, \, 0,
\eeqr
which is a genus-zero curve with the simple rational 
parameterization
\beqr
\label{param}
\hspace{-0.95in}&&  \quad \quad \quad  \quad  \quad \,\,\, \,
C \, \,  =  \, \,  \,
\left({{3 \, x} \over {1\, -x}}   \right)^3, \qquad \, 
D  \, \,  = \, \,  \,
\left({\frac { 1\, -4\,x}{ 1 \, +5\,x }}\right)^3,
\eeqr
such that
\beqr
\hspace{-0.95in}&&  \quad \quad \,\,\,\,\,\,\,
D(x)  \, \,  = \, \,  \, 
C\left( {{1\, -4 \, x} \over {4 \, + \, 11 \, x}} \right),
 \quad \quad \,\,\,\,
C(x)  \, \,  = \, \,  \, 
D\left( {{1\, -4 \, x} \over {4 \, + \, 11 \, x}} \right). 
\eeqr
Along this line, introducing $\, N_2^{p}$ as 
the $\, (1\, -4 \, x)/(4 \, + \, 11 \, x)$ 
pullback of the linear differential operator $\,N_2$, 
one sees that the symmetric square of  $\, N_2^{p}$ 
and of $ \, N_2$ are actually homomorphic. 

With $\, C$ and $\, D$ given by (\ref{param}), and thus 
related by the modular curve (\ref{modular}), one has the 
following non-trivial identity on the {\em same} $\, _2F_1$ 
hypergeometric function with the two different pullbacks 
$\, 1 \, -D$ and $\, C$ 
\beqr
\label{modularequ}
\hspace{-0.95in}&&  \quad  \quad \quad  \quad 
_2F_1\left(\left[{{1} \over {3}}, {{2} \over {3}}\right], 
\, [1], \, 1\, -D\right) 
\, \, = \, \, \,   \,
{{ 1 \, + 5 \, x} \over { 1 \, -x}}
 \cdot \, 
 _2F_1\left(\left[{{1} \over {3}}, {{2} \over {3}}\right],
 \, [1], \, C\right), 
 \eeqr
namely
\beqr
\hspace{-0.98in}&& 
_2F_1\left(\left[{{1} \over {3}}, {{2} \over {3}}\right], \, [1], \, 
{\frac {27 \, x \cdot \,(1\,+x\, +7\,{x}^{2})}{(1\, +5\,x)^{3}}}\right) 
\,=\, \,  
{{ 1 \, + 5 \, x} \over { 1 \, -x}}
 \cdot \,  _2F_1\left(\left[{{1} \over {3}}, {{2} \over {3}}\right], \, [1], 
\, {\frac {27\,{x}^{3}}{ (1\, -x)^{3}}}\right).
\nonumber 
\eeqr
This relation is (after the change of variable 
$\, x \, \rightarrow \, 3 \, x/(1 \, -x)$,
 nothing but {\em  Ramanujan's Cubic transformation}
(see Cor. 2.4 page 97 of~\cite{RamanujanNote}
and (2.23) in~\cite{Ramanujan})
\beqr
\hspace{-0.95in}&& \, \, \,\,
(1\, +2\, x) \cdot \,
 _2F_1\left(\left[{{1} \over {3}}, {{2} \over {3}}\right], 
\, [1], \, x^3 \right) 
\,\,=\, \, \, 
_2F_1\left(\left[{{1} \over {3}}, {{2} \over {3}}\right], \, [1], \, 
1 \, - \, \Bigl( {{ 1\, -x} \over { 1\, + 2 \, x}}  \Bigr)^3
\right).
\eeqr

This relation is also, up to a simple change of variables,
the relation on page 44, Table 18, fifth line 
in Maier's paper~\cite{maier2009}
\beqr
\label{modularequ2}
\hspace{-0.95in}&&  \quad \quad \quad \quad  \,  \,  \,  \, 
_2F_1\left(\left[{{1} \over {3}}, {{2} \over {3}}\right], \, [1], \, 
{\frac {x \cdot \, ({x}^{2}+9\,x+27) }{ (x \,+3)^{3}}}\right) 
\nonumber \\ 
\hspace{-0.95in}&& \quad \quad \quad  \quad 
\quad  \quad \quad \quad  \,  \, 
 \,= \, \,\,\,
3 \cdot \, {{ x \, + \,3} \over { x \, + \, 9}}
 \cdot \,  _2F_1\left(\left[{{1} \over {3}}, {{2} \over {3}}\right], \, [1], \,
{\frac {{x}^{3}}{ (x \,+9)^{3}}}\right).
\eeqr
Such non-trivial identities on the {\em same} $\, _2F_1$ 
hypergeometric function with the {\em two different pullbacks} 
related by a {\em modular curve}, show the emergence of 
a {\em modular form} (see Maier's paper~\cite{maier2009}). 


\subsection{The solutions of $\, N_3$ as modular forms}
\label{ident}

The solution of $ \, N_3$  is given in terms of the solution of 
$ \, V_2$ given in either the form  (\ref{solN3}) or (\ref{othersolN3}),
 through (\ref{N3toU2hom}). 
The fact that the same solution series (\ref{sersolV2}) 
can be expressed {\em in two different ways}, (\ref{solN3}) 
or (\ref{othersolN3}), 
corresponds to a {\em  quite non-trivial identity}, namely  
\beqr
\label{twopullnontrivial}
\hspace{-0.95in}&&  \quad \quad \quad  \quad   \quad \quad 
 {\cal S}(x, \, U) \,   \, = \,  \, \, \,
 -5^{-1/2} \cdot \, {\cal S}(x, \, -U),
\eeqr
between the { same hypergeometric function but with two 
different algebraic pullbacks}.
Such non-trivial identity {\em actually corresponds to a modular form}
(a covariance with respect to the isogenies associated 
with the {\em modular curve}~\cite{bostan2013bchm,Modular}).

These two {\em algebraic} pullbacks 
\beqr
\label{twopull}
\hspace{-0.95in}&&  \quad \quad \quad  \, \,  \quad   \, \,
A \, \, = \, \, \, {\frac {4096\,{x}^{10}}{ (1 -4\,{x}^{2} \, -U)^{4}}}, 
\qquad \, 
B \, \, = \, \, \, {\frac {4096\,{x}^{10}}{ (1 -4\,{x}^{2} \, +U)^{4}}}, 
\eeqr
are related by the {\em genus-zero} modular curve
\begin{small}
\beqr
\label{modularcurve}
\hspace{-0.95in}&&  
 \,\, 722204136308736 \cdot \,  A^4 B^4 \cdot \,  (625A^2\,+1054AB\,+625B^2)
\nonumber\\
\hspace{-0.95in}&&  \quad \, 
+13931406950400 \cdot \,  A^3 B^3 \cdot \,  
(A+B) \cdot \,  (5A^2\,-43786\,AB\,+5B^2)
\nonumber\\
\hspace{-0.95in}&&  \quad \, \,
+1343692800 \cdot \,  A^2B^2 \cdot \,  
[3\, (A^4+B^4)-15066308\, AB \cdot \, (A^2+B^2) +114938242\, A^2B^2]
\nonumber\\
\hspace{-0.95in}&&  \quad \, \,
+103680 \cdot \, AB \cdot \, (A+B) \cdot \, 
[A^4+B^4 +45004444 \, AB \cdot \, (A^2+B^2) \, +35527135712\,6A^2B^2]
\nonumber\\
\hspace{-0.95in}&&  \quad \, \,
+204800\, AB \cdot \, (A+B)\, (6137 A^2\, +847562510 \, AB\, +6137 B^2)
\nonumber\\
\hspace{-0.95in}&&  \quad \, \,
-6553600\cdot \,  AB \cdot \,  (863A^2-3718702AB+863B^2)
\, \, +8724152320\cdot \, AB \cdot \, (A+B)
\nonumber\\
\hspace{-0.95in}&&  \quad \, \,
+[A^6+B^6 \, -65094150\, AB\cdot \, (A^4+B^4)\,
 -13453926179834900 \cdot \, A^3B^3
\nonumber\\
\hspace{-0.95in}&&  \quad \quad \quad  \quad  \quad 
\label{ABmodcurve}
\,\, +98471158056975\, A^2B^2\cdot \, (A^2+B^2)]
\nonumber\\
\hspace{-0.95in}&&  \quad \, \,
 -4294967296 \cdot \, AB \,\,\, \, = \,\, \, \, 0. 
\eeqr
\end{small}

This genus-zero modular curve can be seen 
as corresponding to the elimination of the $\, x$ variable 
between the two ``auxilliary equations'' 
(see (70) in~\cite{bostan2013bchm})
\begin{eqnarray}
\label{Aauxeq}
\hspace{-0.95in}&&   \quad  \quad  
(1\, +20\,x^{2})^{4}  \cdot \, A^{2} \,  \,  \, 
+256 \,{x}^{2} \cdot \, 
(224\,{x}^{8}-400\,{x}^{6}-50\,{x}^{4}+20\,{x}^{2}-1)  \cdot \,  A
\nonumber \\ 
\hspace{-0.95in}&&  \quad  \quad  \quad  \quad  \quad  \quad  \quad 
\, +65536 \cdot \,{x}^{12} \,\, =\, \,\,\,0, 
\\
\hspace{-0.95in}&&  \quad \quad  
\label{Bauxeq}
(1\, +20\,x^{2})^{4}  \cdot \, B^{2} \,  \,  \, 
+256 \,{x}^{2} \cdot \, 
(224\,{x}^{8}-400\,{x}^{6}-50\,{x}^{4}+20\,{x}^{2}-1)  \cdot \,  B
\nonumber \\ 
\hspace{-0.95in}&&  \quad  \quad  \quad  \quad  \quad  \quad  \quad 
\, +65536 \cdot \,{x}^{12} \,\, =\, \,\,\,  0.
\end{eqnarray}
These two auxiliary equations are actually, and surprisingly, 
{\em genus-one} curves rather than genus zero: this is a consequence 
of the fact that the exact expression 
of the two algebraic pullbacks in (\ref{Aauxeq}) 
and (\ref{Bauxeq}) requires a square root, namely $\,U$. On 
the other hand, the modular curve 
 (\ref{ABmodcurve}) that one would expect to be a genus-one curve 
{\em is in fact a genus-zero curve}.

\vskip .1cm 

We note that the pullback of the $ \,\, _2F_1([1/8,3/8],[1], \,z)\,$ 
hypergeometric function in the solution of $ \,N_3$ 
can be rewritten as
\beqr
\label{rewritten}
{\frac {4096\,{x}^{10}}{ (1 -4\,{x}^{2} \, -U)^{4}}}
 \, \,   = \, \,      \,  
16 \, \, x^2 \cdot \, 
\left({{1 -4\,{x}^{2} \, +U } \over {1 \, +20\, x^2 }} \right)^4
\eeqr
which can be rewritten in an alternative way which is linear in $ \,U$
\beqr
\label{rewritten2}
\hspace{-0.95in}&&  \quad  \quad  \quad    \quad 
{\frac {128~{x}^{2} \cdot \, (
 1\, -20\,{x}^{2}\, +50\,{x}^{4}\, +400\,{x}^{6} \,
-224\,{x}^{8}) }{ (1\, + 20\,{x}^{2})^{4}}}
 \nonumber\\
\hspace{-0.95in}&&  \qquad \quad \qquad \quad    \quad 
+ \,  U \cdot \, 
{\frac {128~{x}^{2} \cdot \, (1 \, -4 \,x^2)  \, 
(1\, + 2\,{x}^{2})  \, (1\,  -12\,{x}^{2}) }{ (1\, + 20\,{x}^{2})^{4}}}.
\eeqr
From this rewriting of the pullback, it is tempting to see the singularity 
$ \,1 \,+20x^2 \, = \,0$ of the pullback as a singularity of 
the function. This is not the case, as can be seen in \ref{app:pullback}.

\vskip .2cm 

\subsubsection{Another parametrization \\}
\label{anotherparam}

If one recalls the definition of the square root variable $\, U$ 
in (\ref{Udef}) and the previous expressions for the 
pullback $\, A$,  in (\ref{rewritten}), one 
remarks that all these expressions are, in fact,
functions of $\, X \, = \, \, x^2$. The definition (\ref{Udef})
of $\, U$ corresponds to a {\em rational curve} 
$\,\,U^2 \,\,-(1\, -16\,X)\,(1  \, +4\,X) \, = \, \, 0$, which 
can be parametrized as follows
\beqr
\hspace{-0.95in}&&  \qquad \quad  \quad 
U \, = \, \,  {{5} \over { 4}} \cdot  \, 
{\frac {{t}^{2} \, -64}{{t}^{2} \, +64}}, 
\qquad \quad 
X \, = \, \, 
-{{1} \over {32}} \cdot  \, 
{\frac { (3\,t-8)  \, (t-24) }{{t}^{2} \, +64}}, 
\eeqr
yielding the following rational parametrization of $\, A$ and $\, B$ 
for the genus-zero modular curve (\ref{modularcurve})
\beqr
\label{ABparam}
\hspace{-0.95in}&& 
A(t) =  \, 
- {{1} \over {2}} 
{\frac { (3\,t \, -8)  \, (t \, -24)^{5}}{
 (7\,t \, -8)^{4} \, ({t}^{2} \,+64) }}, 
 \,   \quad 
B(t)  = \,  
- {{1} \over {2}} 
{\frac { (3\,t \, -8)^{5} \,  (t \, -24) }{
 (t \, -56)^{4} \, ({t}^{2} \, +64) }}
  \,  =  \,  \, A\left({{64} \over {t}} \right).
\eeqr
Performing the change of variable $\, t \, = \, 24 \, +u$, 
one has the alternative parametrization
\beqr
\hspace{-0.95in}&&  \quad \quad  \, 
U \, = \, \, 
{{5} \over {4}} \cdot \,{\frac { (u \, +32)  \, (u \, +16) }{
{u}^{2}+48\,u+640}},
\qquad \quad 
X \, = \, \, 
- {{1} \over {32}} \cdot \,{\frac { 
(64+3\,u)\cdot \,  u}{{u}^{2}+48\,u+640}}, 
\eeqr
\beqr
\label{ABparam}
\hspace{-0.95in}&& 
A(u) =  \, -{{1} \over {2}}
\,{\frac { (64+3\,u) \cdot \,  {u}^{5}}{
 (160+7\,u)^{4} \, ({u}^{2}+48\,u+640) }}, 
 \,   \quad 
B(u)  = \,  - {{1} \over {2}} 
\,{\frac {(64 \, +3\,u)^{5} \cdot \, u}{ 
(u-32)^{4} \, ({u}^{2}+48\,u+640) }}.
\nonumber 
\eeqr
Rewriting the solution of $\, N_3$ in terms of the $\, _3F_2$ 
hypergeometric function (\ref{pullbacked3F2})  amounts to 
considering the following $\, _3F_2$  identity
\beqr
\label{identity3F2}
\hspace{-0.95in}&& \qquad \quad \, \,
(160 \, -5\, u) \cdot \,
 _3F_2\left(\left[{{1} \over {4}}, \, {{1} \over {2}}, \,  {{3} \over {4}}\right],
 \, [1, \, 1], \, \, A(u)\right)
\nonumber \\ 
\hspace{-0.95in}&&  \qquad \quad \quad \quad \quad \quad \quad 
\,   = \, \,  \, 
(160 \, + 7 \, u) \cdot \, 
_3F_2\left(\left[{{1} \over {4}}, \, {{1} \over {2}}, \,  {{3} \over {4}}\right],
 \, [1, \, 1], \, \, B(u)\right),
\eeqr
which corresponds, using the Clausen identity (\ref{Clausen}), to 
the $\, _2F_1$  identity
\beqr
\label{identity2F1}
\hspace{-0.95in}&& \qquad \quad  \, \,
(160 \, -5\, u)^{1/2}  \cdot \, 
_2F_1\left(\left[{{1} \over {8}}, \,  {{3} \over {8}}\right], \, [1], \, A(u) \right)
\nonumber \\ 
\hspace{-0.95in}&&  \qquad \quad \quad \quad \quad \quad  \quad  \quad 
\,   = \, \,  \, (160 \, + 7 \, u)^{1/2}  \cdot \,
_2F_1\left(\left[{{1} \over {8}}, \,  {{3} \over {8}}\right], \, [1], \, B(u) \right).
\eeqr

\subsubsection{Infinite order symmetry on a Heun function \\}
\label{infiniteHeunG}

The occurrence of modular forms corresponds to identities like 
(\ref{twopullnontrivial}) or (\ref{identity2F1}), relating the same 
$\, _2F_1$  hypergeometric function with two different pullbacks, 
which are related by a  modular curve (\ref{modularcurve}). 
These infinite order symmetries of the $\, _2F_1$  hypergeometric 
functions corresponds to {\em isogenies}~\cite{bostan2013bchm,Modular} 
{\em of the elliptic curves} which amount to multiplying 
or dividing the ratio of the two periods of an elliptic 
curve by an integer $\, N$.

The solution of $\,\mathrm{Sol}(V_2)$, expressed in terms of 
a $\, _2F_1$  hypergeometric function (\ref{solN3}), can also be expressed
as a simple Heun function (\ref{HeunGsol}). One can thus expect 
an {\em infinite order symmetry} identity on this  Heun 
function. The identity reads
\begin{eqnarray}
\label{identHeunG}
\hspace{-0.95in}&& \quad \quad \quad \quad \quad 
 {\cal A}_1(X) \cdot \, 
\mathrm{Heun}\Bigl( -{{1} \over {4}}, \, {{1} \over {16}}, \, \,
 {{3} \over {8}}, \, {{5} \over {8}},\, 1, \, {{1} \over {2}}, \,  -4 \, X\Bigr)
\nonumber\\
\hspace{-0.95in}&&  \, \quad \quad \quad \quad \quad \quad \quad 
= \, \,\,\,  {\cal A}_2(Y) \cdot \,
\mathrm{Heun}\Bigl( -{{1} \over {4}}, \, {{1} \over {16}}, \, \,
 {{3} \over {8}}, \, {{5} \over {8}},\, 1, \, {{1} \over {2}}, \,  -4 \, Y\Bigr), 
\end{eqnarray}
where $\, X$ and $\, Y$ are related by a {\em genus-one} curve 
$\, P(X, \, Y) \, = \, \, 0$ given in \ref{infinite},
and $\, {\cal A}_1(X)$ and $\, {\cal A}_2(X)$ are two algebraic expressions
also given  in \ref{infinite}. If the expression of the solution 
of $\,\mathrm{Sol}(V_2)$ in terms of Heun function looks (artificially)
simpler, the representation of the infinite order isogeny symmetries  
is more involved, since we do not have a rational parametrization
of $\, P(X, \, Y) \, = \, \, 0$.


\subsection{$N_2$ versus $\, N_3$}
\label{N2versusN3}

We can attempt to find a relationship between the ${}_2F_1$ hypergeometric 
functions appearing in the solutions of $ \,N_2$ and $ \,N_3$ (via $ \,V_2$). 
In order to achieve that goal let us rather try to reduce both their 
corresponding $\, _2F_1$ hypergeometric functions with a pullback, 
$ \,_2F_1([1/3,2/3],[1], \, r(x))$ and $ \, _2F_1([1/8,3/8],[1],\, s(x))$,  
to a standard~\cite{maier2009}
$ \,_2F_1([1/12,5/12],[1], \, t(x))$ form. This can indeed be done, 
according to the two identities below
\beqr
\label{rewritten2}
\hspace*{-\mathindent}&& \quad  \quad 
_2F_1\left(\left[\frac{1}{3},\frac{2}{3}\right],[1], \,x\right)  \, \,
= \, \,  \, 
\frac{1}{Q^{1/4}} \cdot \,
 _2F_1\left(\left[\frac{1}{12},\frac{5}{12}\right],[1],
 \, -\frac{64x\cdot(x-1)^3}{(8x+1)^3}\right), 
\eeqr
where
\beqr
\label{Q54}
\hspace{-0.95in}&&  \qquad \quad \quad  \quad  \qquad \,\,
Q \, \, = \, \, \, 5 \,  \, - 4\,\sqrt{1\, -4 \,x \cdot \, (1-x)},
\eeqr
and
\beqr
\label{rewritten2bis}
\hspace{-0.95in}&& \quad \quad \quad  \quad  \, \,
 _2F_1\left(\left[\frac{1}{8},\frac{3}{8}\right],[1], 
\,-\frac{4x}{(1-x)^2}\right) 
\nonumber \\
\hspace{-0.95in}&& \qquad \quad \qquad \quad
 \, \, \, \,
=  \, \, \, \left( \frac{1\, -x}{1\, -4x} \right)^{1/4}
\cdot \, {}_2F_1\left(\left[\frac{1}{12},\frac{5}{12}\right],[1], \,
-\frac{27x}{(1-4x)^3} \right).
\eeqr

Using the first identity (\ref{rewritten2}) on 
the $\, _2F_1$ hypergeometric function (\ref{N2solH}), 
occurring in the solution (\ref{N2sol}) of the second order
operator $\, N_2$ yields
\beqr
\label{Q54next}
\hspace*{-\mathindent}
&&\quad \quad 
_2F_1\left(\left[\frac{1}{3},\frac{2}{3}\right],[1],
 \, \frac{27x^3}{(1-x)^3}\right)
\nonumber\\
\hspace*{-\mathindent} &&\quad \quad \quad 
 \,   \,\, =\, \,\,   \, \, \frac{(1\,-x)^{3/4} }{
(1\, +5 \, x)^{1/4} \cdot \,
 (1\, \, -8 \, x \, + \, 43 \, x^2  )^{1/4}} \cdot \, 
{}_2F_1\left(\left[\frac{1}{12},\frac{5}{12} \right],[1],
\, \, {\cal P}_2\right), 
\eeqr
where the pullback $\, {\cal P}_2$ reads
\beqr
\label{callP2}
\hspace{-0.95in}&& \quad \quad \quad \quad \quad  \quad
 {\cal P}_2 \, \, = \, \, \, 
\frac{1728 \, \, x^3 \cdot \, (1 \, -4 \, x)^3 
\, (1 \, +\, x \, +7 \, x^2)^3  }{
(1 \, -x)^3 \, (1\, +5 \, x)^3 \, (1\, -8 \, x \, + \, 43 \, x^2)^3}.
\eeqr
Similarly, using the second identity (\ref{rewritten2bis}) on 
the solution $\, \mathrm{Sol}(V_2)$ given by (\ref{solN3})
occurring in the solution of the 3rd order operator 
$\, N_3$ yields a rewriting of $\, \mathrm{Sol}(V_2)$ 
in terms of a pullbacked hypergeometric function
$ \,_2F_1([1/12,5/12],[1], \, {\cal P}_3)$. 

The elimination of $\, x$ between these two pullbacks $\, {\cal P}_2$
and  $\, {\cal P}_3$ yields an involved polynomial relation 
$\, P({\cal P}_2, \, {\cal P}_3) \, \, = \, \, \, 0$, where the polynomial
$\, P$ is the sum
of $\,1665$ monomials of degree  $\, 36$ in $\, {\cal P}_2$ and  
$\, 48$ in $\, {\cal P}_3$. In other words, the solutions of 
$\, \mathrm{Sol}(N_2)$ and $\mathrm{Sol}(N_3)$ and their corresponding
modular forms are far 
from being simply related.


\section{Towards generalizations of the results}
\label{sec:discussion}

There are several ways in which the results of this paper could be extended. 
From~\cite{rechnitzer2003}, it is known that the {\em anisotropic perimeter} 
generating functions for three-choice and imperfect staircase polygons 
have a simple structure. Since for all known closed-form solutions it has 
been shown~\cite{rechnitzer2000} that the anisotropic perimeter generating 
functions are simple extensions of their isotropic counterparts, one could 
expect that the anisotropic versions of the generating functions in this 
paper could be simple extensions of the hypergeometric functions appearing 
in the solutions. It may be that only the arguments (pullbacks) of the 
$\, {}_2F_1$ hypergeometric functions become two-variable rational or 
algebraic functions.  Another plausible extension would 
be generalizations to two-variable hypergeometric functions, such 
as Appell or Horn functions~\cite{bateman1953}.

A second generalization of the results would be to consider the 
{\em area-perimeter} generating function. All known results for 
area-perimeter generating functions involve $\, q$-series~\cite{rechnitzer2000}. 
In~\cite{guttmann2006j} and~\cite{guttmann2006j2} conjectured forms 
for the area-perimeter generating functions are proposed for 
three-choice and imperfect staircase polygons, 
and 1-punctured staircase polygons, 
respectively. The conjectures involve $\, q$-Bessel functions with 
algebraic pre-factors. Alternatively, based on the hypergeometric 
results above, it is reasonable to propose the 
appearance of $\, q$-hypergeometric 
functions, also called basic hypergeometric functions~\cite{Gasper}. We note 
that they have already appeared in the SAP area generating function 
of prudent polygons in~\cite{beaton2011fg}.

Finally, it is possible to consider the effect of increasing the number of 
punctures for punctured staircase polygons. In~\cite{guttmann2000jwe}, 
the effect of increasing the number of punctures was considered:  it 
was found that as the number of punctures increases, 
the perimeter generating function critical exponent increases by 3/2 
per puncture, while the area generating function critical exponent 
increases by 1 per puncture. In both cases, the critical point was found 
to be unchanged by a finite number of punctures. However, 
in~\cite{guttmann2001jo}, it was found that once the number of punctures is 
allowed to be unbounded, the perimeter generating function has a zero radius of 
convergence. Considering our $\, {}_2F_1$ hypergeometric function 
representation of the 1-punctured perimeter generating function, a simple scenario 
that could explain these properties, would be that going from one to $\, n$ 
punctures, the $\, {}_2F_1$ hypergeometric functions are of the form 
$\, {}_2F_1([a \, +3\, n/2,\, b],[c],\, {\cal P}(x))$. Under 
this scenario, for finite $\, n$, there is a critical exponent 
increase of $\, 3/2 \, $ per puncture, while the critical point 
remains unchanged, and as $\, n\, \to \infty$, corresponding to an unbounded 
number of punctures, the hypergeometric function will become a confluent 
hypergeometric function with the critical point mapping to the confluent 
irregular singularity at infinity, whose series will have zero radius 
of convergence, in agreement with what was found in~\cite{guttmann2001jo}.


\section{Conclusions}
\label{conclusions}

We have demonstrated for the first time a non-algebraic, $\, D$-finite perimeter 
generating function for SAPs, given in terms of $ \, {}_2F_1$ hypergeometric 
functions, and we have provided simple relationships between the generating 
functions of three-choice, imperfect, and 1-punctured staircase polygons. We 
have expressed the generating functions as a sum of algebraic and transcendental 
parts, each of which is a series in integer coefficients up to an overall factor 
of $\, 1/60$. We have been able to fully analyze the solutions of their 8th order 
linear differential operators since they, up to the semi-direct product, reduce 
to a 3rd order, a 2nd order, and first order operators. We 
have found that the 2nd order operator 
has {\em modular form} solutions which can be rewritten as a $\, {}_2F_1$ 
hypergeometric function with {\em two possible pullbacks}. Similarly we have 
found that the 3rd order operator is homomorphic to the symmetric square of 
an 2nd order operator which also has solutions in terms of another 
{\em modular form} which, again, can be expressed as 
a $\, {}_2F_1$ hypergeometric function with 
two possible pullbacks. In that case, these two pullbacks are 
related by a genus-zero modular curve.
These two {\em modular forms} 
are not simply related, as can be seen when 
one rewrites them in terms of a common 
$\, {}_2F_1([1/12,5/12],[1],\, {\cal P}(x)) \, $ 
hypergeometric functions for respective $\,  {\cal P}(x)$  pullbacks. 
All these exact results for the three perimeter generating functions illustrate, 
one more time~\cite{bostan2013bchm,Modular}, the emergence in enumerative 
combinatorics and lattice statistical mechanics
of (quite non-trivial) {\em modular forms}. The emergence of modular forms
is often a consequence of the fact that the functions one considers 
in enumerative combinatorics and lattice statistical mechanics, can also 
be written as $\, n$-fold integrals and are, in fact, diagonal of rational 
functions~\cite{bostan2013bchm,Diago2}. One can reasonably conjecture that the
generating functions analysed here are actually diagonal of rational 
functions.

\ack 
This work has been performed without
 any support of the ANR, the ERC, the MAE or any PES of the CNRS.
One of us (M.A.) would like to thank the Australian Research Council
 for supporting this work under the Discovery Project scheme 
(project number DP140101110). We thank Iwan Jensen for providing
 series expansions and the linear ODEs used in this paper. 

\vskip .3cm 

\vskip .3cm 

\vskip .3cm

\appendix


\section{Operator polynomial definitions}
\label{app:operators}

\begin{small}
\beqr
\hspace*{-\mathindent}p_6 &=& 6874x^6-2913x^5+660x^4-230x^3+60x^2+6x-2, 
 \\
\hspace*{-\mathindent}p_7 &=& 13748x^7-1341x^6 +1047x^5
-1510x^4 +600x^3 +42x^2 -34x+3, 
 \\
\hspace*{-\mathindent}p_9 &=& 96236x^9+35756x^8+29198x^7-30049x^6 +10841x^5
-9226x^4+2819x^3-913x^2
\nonumber\\
\hspace*{-\mathindent}&&+224x-21, 
\eeqr
\end{small}
\begin{small}
\beqr
\hspace*{-\mathindent} p_{10} &=& 577416x^{10}+11494x^9-265110x^8
-104347x^7 +14641x^6 -17865x^5 +11006x^4 -2990x^3
\nonumber\\
\hspace*{-\mathindent}&& +582x^2 -46x -6, 
 \\
\hspace*{-\mathindent}p_{14} &=& 258682368x^{14} +5149312x^{13}
-116080384x^{12}-56911264x^{11}-11623368x^{10} -16910078x^9
\nonumber\\
\hspace*{-\mathindent}&& +7172550x^8 -2518103x^7 +1020461x^6
-167793x^5+12300x^4+3408x^3-2168x^2
\nonumber\\
\hspace*{-\mathindent}&& +118x+16,  
\eeqr
\end{small}
\begin{small}
\beqr
\hspace*{-\mathindent}\tilde{p}_{14} &=&1459419136\,{x}^{14}
+1247508864\,{x}^{13}+811733344\,{x}^{12} 
\nonumber\\
\hspace*{-\mathindent}&&+477640288\,{x}^{11}
+138848672\,{x}^{10}+13410136\,{x}^{9} 
\nonumber\\
\hspace*{-\mathindent}&&-3835374\,{x}^{8}
-5661538\,{x}^{7}-1091480\,{x}^{6}+190073\,{x}^{5}
 \nonumber\\
\hspace*{-\mathindent}&&+27848\,{x}^{4}
-13725\,{x}^{3}-7006\,{x}^{2}-708\,x-30,
\\
\hspace*{-\mathindent}\tilde{q}_{14} &=&25467129856\,{x}^{14}
+23296248192\,{x}^{13}+15147763936\,{x}^{12}
 \nonumber\\
\hspace*{-\mathindent}&&+9747680992\,{x}^{11}
+2515397616\,{x}^{10}+986365576\,{x}^{9} 
\nonumber\\
\hspace*{-\mathindent}&&-122753886\,{x}^{8}
-50126514\,{x}^{7}-16282341\,{x}^{6} 
\nonumber\\
\hspace*{-\mathindent}&&-5516044\,{x}^{5}
+1401243\,{x}^{4}+113789\,{x}^{3}-83497\,{x}^{2}-4836\,x-82, 
\\
\hspace*{-\mathindent}\tilde{p}_{15} &=&3201028096\,{x}^{15}
+3149819904\,{x}^{14}+2205543168\,{x}^{13}+1545006784\,{x}^{12} 
\nonumber\\
\hspace*{-\mathindent}&&
+432099808\,{x}^{11}+194591088\,{x}^{10}
-26426336\,{x}^{9}-11116244\,{x}^{8}-4340246\,{x}^{7}
 \nonumber\\
\hspace*{-\mathindent}&&
-1902039\,{x}^{6}+549358\,{x}^{5}+63757\,{x}^{4}-50994\,{x}^{3} 
\nonumber\\
\hspace*{-\mathindent}&&
-4128\,{x}^{2}-104\,x+3,
\eeqr
\end{small}
\begin{small}
\beqr
\hspace*{-\mathindent}p_{18} &=& 9312565248x^{18}+185375232x^{17}
-4282264448x^{16}-2497630752x^{15}-973632640x^{14}
\nonumber\\
\hspace*{-\mathindent}&& -1001053992x^{13}+183496040x^{12}-202323868x^{11}
+81320436x^{10}-20237208x^9+7144232x^8
\nonumber\\
\hspace*{-\mathindent}&& -1005079x^7-46763x^6+31581x^5
 -13467x^4 -853x^3 +717x^2 -36x -5,  
\\
\hspace*{-\mathindent}p_{23} &=& 22984790671360x^{23}
-14160990742528x^{22}
+47196432034304x^{21}-40184041956352x^{20}
\nonumber\\
\hspace*{-\mathindent}&& +23871790843776x^{19}-12862188171584x^{18}
+4334321680992x^{17}-808339934032x^{16}
\nonumber\\
\hspace*{-\mathindent}&& +70414369000x^{15}+59364489644x^{14}
-38533903096x^{13}+4418397469x^{12}+2623726024x^{11}
\nonumber\\
\hspace*{-\mathindent}&& -1386913106x^{10}+512965024x^{9}
-144171921x^{8}+17788918x^{7}+2272607x^{6}-949665x^{5}
\nonumber\\
\hspace*{-\mathindent}&& +63356x^{4}+6516x^{3}-426x^{2}-28x-2,
 \eeqr
\end{small}
\begin{small}
\beqr
\hspace*{-\mathindent}p_{24} &=& 1103269952225280x^{24}
-228305326678016x^{23}+2331485726244864x^{22}
\nonumber\\
\hspace*{-\mathindent}&& -1249271454269440x^{21}
+549381083516928x^{20}-225290952722816x^{19}
\nonumber\\\hspace*{-\mathindent}&& -39003496295360x^{18}
+46746500840896x^{17} -10249554621312x^{16}+1973847887848x^{15}
\nonumber\\\hspace*{-\mathindent}&& +157900491180x^{14}
-637108022672x^{13} +233984558200x^{12}-24645390372x^{11}
\nonumber\\
\hspace*{-\mathindent}&& -4177273140x^{10}+2621821288x^{9}
-942904492x^{8}+195411966x^{7}+1609130x^{6}
\nonumber\\
\hspace*{-\mathindent}&& -6956791x^{5}+515168x^{4}
+60240x^{3}-2676x^{2}-256x-20, 
 \\
\hspace*{-\mathindent}q_{24} &=& 91939162685440x^{24}
-79628753641472x^{23} +202946718879744x^{22}-207932599859712x^{21}
\nonumber\\
\hspace*{-\mathindent}&& +135671205331456x^{20}-75320543530112x^{19}
+30199474895552x^{18}-7567681417120x^{17}
\nonumber\\
\hspace*{-\mathindent}&& +1089997410032x^{16}+167043589576x^{15}
-213500102028x^{14}+56207492972x^{13}
\nonumber\\
\hspace*{-\mathindent}&& +6076506627x^{12}-8171378448x^{11}
+3438773202x^{10}-1089652708x^{9}+215327593x^{8}
\nonumber\\
\hspace*{-\mathindent}&& -8698490x^{7}-6071267x^{6}
+1203089x^{5}-37292x^{4}-8220x^{3}+314x^{2}+20x+2, 
 \eeqr
\end{small}
\begin{small}
\beqr
\hspace*{-\mathindent}p_{31} &=& 42659311790230732800x^{31}
-29563777128137269248x^{30}+109223631121278908416x^{29}
\nonumber\\
\hspace*{-\mathindent}&& -104250789052923003904x^{28}
+56305599211721642496x^{27}-24632274570479903488x^{26}
\nonumber\\
\hspace*{-\mathindent}&& +2681522191975403520x^{25}
+4954045657465530112x^{24} -3002212360825142752x^{23}
\nonumber\\
\hspace*{-\mathindent}&& +813437846722409936x^{22}
-111297129389473336x^{21} -52910191450930076x^{20}
\nonumber\\
\hspace*{-\mathindent}&& +54606842707567716x^{19}
-20972722051528144x^{18}+4010775763371668x^{17}
\nonumber\\
\hspace*{-\mathindent}&& -56829134814870x^{16}
-339637124743736x^{15} +184268988549260x^{14}
\nonumber\\
\hspace*{-\mathindent}&& -53836053904996x^{13}
+8015086193990x^{12} -65165405654x^{11}-228234905736x^{10}
\nonumber\\
\hspace*{-\mathindent}&& +60861828179x^{9}-9657526198x^{8}
+348474792x^{7}+171633727x^{6}-20489803x^{5}
\nonumber\\
\hspace*{-\mathindent}&& -752308x^{4}+120494x^{3}+5118x^{2}+196x-46,
\eeqr
\end{small}
\begin{footnotesize}
\beqr
\hspace*{-\mathindent}p_{33} &=& 598165554871664640\,{x}^{33}
+2237833352545566720\,{x}^{32}
+2923793972548599808\,{x}^{31}
\nonumber\\
\hspace*{-\mathindent}&&
+4898487216451354624\,{x}^{30}+3066215578973962240\,{x}^{29}
+3807302765273284608\,{x}^{28}
\nonumber\\
\hspace*{-\mathindent}&&+1548493928510070784\,{x}^{27}
+993756825730510848\,{x}^{26}
+330665809850894336\,{x}^{25}
\nonumber\\
\hspace*{-\mathindent}&&+30479967060547584\,{x}^{24}
-10206189043353856\,{x}^{23}
-12653592201109760\,{x}^{22}
\nonumber\\
\hspace*{-\mathindent}&&-11185832980157184\,{x}^{21}
+210438316943104\,{x}^{20}
+61265169248832\,{x}^{19}
\nonumber\\
\hspace*{-\mathindent}&&+26357534470800\,{x}^{18}
+181882051733304\,{x}^{17}
-22437164475672\,{x}^{16}
\nonumber\\
\hspace*{-\mathindent}&&-13063730138481\,{x}^{15}
+1615865720985\,{x}^{14}
-1270457215869\,{x}^{13}+262962192538\,{x}^{12}
\nonumber\\
\hspace*{-\mathindent}&&+171533661840\,{x}^{11}
-44224719936\,{x}^{10}-5369927527\,{x}^{9}+2555317932\,{x}^{8}
\nonumber\\
\hspace*{-\mathindent}&&+150451837\,{x}^{7}
-51686841\,{x}^{6}-5506775\,{x}^{5}
\nonumber\\
\hspace*{-\mathindent}&&
+562261\,{x}^{4}+119245\,{x}^{3}+6577\,{x}^{2}+201\,x+3, 
\eeqr
\end{footnotesize}
\begin{footnotesize}
\beqr
\hspace*{-\mathindent}p_{36} &=& 1186760460865382645760\,{x}^{36}
+4296640036887488102400\,{x}^{35}
+5656257186120920465408\,{x}^{34}
\nonumber\\
\hspace*{-\mathindent}&&
+9610563099027778306048\,{x}^{33}
+6331246887273737748480\,{x}^{32}+7968414685458358861824\,{x}^{31}
\nonumber\\
\hspace*{-\mathindent}&&
+3417201107002357972992\,{x}^{30}
+2613021963777068236800\,{x}^{29}+880949926005413642240\,{x}^{28}
\nonumber\\
\hspace*{-\mathindent}&&
+234438363912752283648\,{x}^{27}
+39705299093018075136\,{x}^{26}-33621804577702641664\,{x}^{25}
\nonumber\\
\hspace*{-\mathindent}&&
-24563052057588912128\,{x}^{24}
-3677799503014345728\,{x}^{23}-2383201063097856256\,{x}^{22}
\nonumber\\
\hspace*{-\mathindent}&&
+353349079985541632\,{x}^{21}
+422742538805020416\,{x}^{20}-18935528677020992\,{x}^{19}
\nonumber\\
\hspace*{-\mathindent}&&+22888292241850368\,{x}^{18}
-3272140352378880\,{x}^{17}-6181103422702752\,{x}^{16}
\nonumber\\
\hspace*{-\mathindent}&&+779365487308732\,{x}^{15}
+111813502211919\,{x}^{14}-10554167286006\,{x}^{13}
\nonumber\\
\hspace*{-\mathindent}&&+34694255695001\,{x}^{12}
-6950573977656\,{x}^{11}
-2354238734992\,{x}^{10}+643662074352\,{x}^{9}
\nonumber\\
\hspace*{-\mathindent}&&
+58104579207\,{x}^{8}-23471784508\,{x}^{7}
-1309856379\,{x}^{6}+330362442\,{x}^{5}+30382891\,{x}^{4}
\nonumber\\
\hspace*{-\mathindent}&&
-2170504\,{x}^{3}
-357735\,{x}^{2}-13190\,x-201, 
\eeqr
\end{footnotesize}
\begin{footnotesize}
\beqr
\hspace*{-\mathindent}q_{36} &=& 4898975894398933401600\,{x}^{36}
+17161731218095095152640\,{x}^{35}
+21883875427370219339776\,{x}^{34}
\nonumber\\
\hspace*{-\mathindent}&&
+35975336852527465365504\,{x}^{33}
+23042462618293310717952\,{x}^{32}+27819843458704542793728\,{x}^{31}
\nonumber\\
\hspace*{-\mathindent}&&+11698815400453787467776\,{x}^{30}
+8612877073871460311040\,{x}^{29}
\nonumber\\
\hspace*{-\mathindent}&&
+2839184277919494885376\,{x}^{28}+762150624816330670080\,{x}^{27}
+129975590943822506496\,{x}^{26}
\nonumber\\
\hspace*{-\mathindent}&&
-89860876746914479616\,{x}^{25}-66319997468290851840\,{x}^{24}
-10529561536853716224\,{x}^{23}
\nonumber\\
\hspace*{-\mathindent}&&-6419230746695990912\,{x}^{22}
+860563513487440608\,{x}^{21}
+974313641274336048\,{x}^{20}
\nonumber\\
\hspace*{-\mathindent}&&
+10482129303048704\,{x}^{19}+70108534257870090\,{x}^{18}
-6926151459159618\,{x}^{17}
\nonumber\\
\hspace*{-\mathindent}&&
-11166062802273588\,{x}^{16}+850683965895387\,{x}^{15}
-39984544775712\,{x}^{14}
\nonumber\\
\hspace*{-\mathindent}&&
+21618692208399\,{x}^{13}+63135026727396\,{x}^{12}
-10612700119674\,{x}^{11}
-3345898146854\,{x}^{10}
\nonumber\\
\hspace*{-\mathindent}&&
+785954523741\,{x}^{9}+58318133424\,{x}^{8}-25700455352\,{x}^{7}
-726830868\,{x}^{6}+322300149\,{x}^{5}
\nonumber\\
\hspace*{-\mathindent}&&+17524916\,{x}^{4}
-1750194\,{x}^{3}-160296\,{x}^{2}-3802\,x-30,
\eeqr
\end{footnotesize}
\begin{footnotesize}
\beqr
\hspace*{-\mathindent}p_{37} &=& 136845384314821493391360x^{37}
-105679883821940306952192x^{36}
+391986252374413297836032x^{35}
\nonumber\\
\hspace*{-\mathindent}&& -399101140153885695805440x^{34}
 +264914197644249574493184x^{33}-202222045944023129525760x^{32}
 \nonumber\\
\hspace*{-\mathindent}&&
+114104519102302216106752x^{31} -43668075787265613729792x^{30}
 +25076614970145903635968x^{29}
 \nonumber\\
\hspace*{-\mathindent}&&
-19991047450482347090016x^{28}+12192274657696530720432x^{27}
 -5894005459795198246136x^{26}
\nonumber\\
\hspace*{-\mathindent}&& +2451165042003983275604x^{25}
-794981526560526083280x^{24}
+152372293756401144616x^{23}
\nonumber\\
\hspace*{-\mathindent}&& +8253142081467241688x^{22}
 -20500019699204933934x^{21}+10371074800492484800x^{20}
\nonumber\\
\hspace*{-\mathindent}&&
-3606002718331668490x^{19} +926435638472444976x^{18} 
-166886391470186702x^{17}
\nonumber\\
\hspace*{-\mathindent}&&
+14211777436985726x^{16}+3766364585838030x^{15}
 -2146994035077380x^{14}+493421682837626x^{13}
\nonumber\\
\hspace*{-\mathindent}&&
 -44957551610202x^{12}-4718419878437x^{11}+1904660277428x^{10}
-282375704850x^{9} 
\nonumber\\
\hspace*{-\mathindent}&& +24803569832x^{8}
 +178483388x^{7} -316167306x^{6}+15725362x^{5}
 \nonumber\\
\hspace*{-\mathindent}&&
+663472x^{4}+136032x^{3}-9204x^{2}-1348x+36,
\eeqr
\end{footnotesize}
\begin{footnotesize}
\beqr
\hspace*{-\mathindent}\tilde{p}_{47} &=&-3998285727059353761208074240\,{x}^{47}
-15722996424986502817413857280\,{x}^{46}
 \nonumber\\
\hspace*{-\mathindent}&&
-31700576980264907590149865472\,{x}^{45}
-39762356689085988698475986944\,{x}^{44}
 \nonumber\\
\hspace*{-\mathindent}&&-55567163415288928360984477696\,{x}^{43}
-41156353869401151264302039040\,{x}^{42}
 \nonumber\\
\hspace*{-\mathindent}&&-42641866704927022312400617472\,{x}^{41}
-27840776428837102543349743616\,{x}^{40}
 \nonumber\\
\hspace*{-\mathindent}&&-22217086391538757419222433792\,{x}^{39}
-11397646311634431011432169472\,{x}^{38}
 \nonumber\\
\hspace*{-\mathindent}&&-7531550773181345715811909632\,{x}^{37}
-2273949231661470224954753024\,{x}^{36}
 \nonumber\\
\hspace*{-\mathindent}&&-1003583489914294831018164224\,{x}^{35}
-112898949096715668230602752\,{x}^{34} 
\nonumber\\
\hspace*{-\mathindent}&&+118387654740438541064945664\,{x}^{33}
+38666219711639414816055296\,{x}^{32}
 \nonumber\\
\hspace*{-\mathindent}&&+35934156638162932949531648\,{x}^{31}
+10426340672815711194877952\,{x}^{30} 
\nonumber\\
\hspace*{-\mathindent}&&-320841457970784345042944\,{x}^{29}
+635223055700342083189760\,{x}^{28}
 \nonumber\\
\hspace*{-\mathindent}&&-284232221387920714589632\,{x}^{27}
-188136397643399775095680\,{x}^{26}
 \nonumber\\
\hspace*{-\mathindent}&&+23294038272449049076032\,{x}^{25}
-14528610751236985394880\,{x}^{24}
 \nonumber\\
\hspace*{-\mathindent}&&-1417939882191877865220\,{x}^{23}
+2231318324009318030740\,{x}^{22}
 \nonumber\\
\hspace*{-\mathindent}&&-508751864781622722470\,{x}^{21}
-40457915193136220663\,{x}^{20} \nonumber\\
\hspace*{-\mathindent}&&+64691874376806398204\,{x}^{19}
-504282932465717984\,{x}^{18} \nonumber\\
\hspace*{-\mathindent}&&+2356742622148449180\,{x}^{17}
+458421645788687707\,{x}^{16}
-726881646184360404\,{x}^{15} \nonumber\\
\hspace*{-\mathindent}&&-56925596051693998\,{x}^{14}
+39718155774616062\,{x}^{13}
+1580647902664907\,{x}^{12} \nonumber\\
\hspace*{-\mathindent}&&-759009159310546\,{x}^{11}-4263037174816\,{x}^{10}
+5800765803232\,{x}^{9} \nonumber\\
\hspace*{-\mathindent}&&+2552969286421\,{x}^{8}+470528432724\,{x}^{7}
-42022916673\,{x}^{6}-17893614632\,{x}^{5} \nonumber\\
\hspace*{-\mathindent}&&-1828148626\,{x}^{4}
-117730660\,{x}^{3}-5201553\,{x}^{2}-126210\,x-1407,
\eeqr
\end{footnotesize}
\begin{footnotesize}
\beqr
\hspace*{-\mathindent}\tilde{p}_{52} &=&1704808334604716622698722099200\,{x}^{52}
+7703187796232335576388924866560\,{x}^{51} \nonumber\\
\hspace*{-\mathindent}&&+15051203287950148717200881483776\,{x}^{50}
+25789357301052060716142075838464\,{x}^{49} \nonumber\\
\hspace*{-\mathindent}&&+30432229952970719844369323524096\,{x}^{48}
+32924880520864943337284088889344\,{x}^{47} \nonumber\\
\hspace*{-\mathindent}&&+29089920691856607650312492154880\,{x}^{46}
+21874357365357016911411287162880\,{x}^{45} \nonumber\\
\hspace*{-\mathindent}&&+15037957669463482332952244781056\,{x}^{44}
+8045705673042215501902878081024\,{x}^{43} \nonumber\\
\hspace*{-\mathindent}&&+4124274395028557902150992134144\,{x}^{42}
+1439011157218352953044700758016\,{x}^{41} \nonumber\\
\hspace*{-\mathindent}&&+414992314542845802067902070784\,{x}^{40}
+9956129705337582530914680832\,{x}^{39} \nonumber\\
\hspace*{-\mathindent}&&-62424097514592658119303954432\,{x}^{38}
-38464628200173359285887041536\,{x}^{37} \nonumber\\
\hspace*{-\mathindent}&&-16953886570364810227683016704\,{x}^{36}
-4048322033447287279393472512\,{x}^{35} \nonumber\\
\hspace*{-\mathindent}&&+146828245861856321486917632\,{x}^{34}
+322245452780508350537965568\,{x}^{33} \nonumber\\
\hspace*{-\mathindent}&&+238574353851711761870308352\,{x}^{32}
+59028509758388482339672064\,{x}^{31} \nonumber\\
\hspace*{-\mathindent}&&-8725510258141260028585216\,{x}^{30}
-1772219094830852759766528\,{x}^{29} \nonumber\\
\hspace*{-\mathindent}&&-2264699219731747758352192\,{x}^{28}
-471388272871319130108672\,{x}^{27} \nonumber\\
\hspace*{-\mathindent}&&+318134935596072152074384\,{x}^{26}
+16840856691026459372880\,{x}^{25} \nonumber\\
\hspace*{-\mathindent}&&+5175955882237784325436\,{x}^{24}
+2983347697005377806624\,{x}^{23} \nonumber\\
\hspace*{-\mathindent}&&-4045729633170366824594\,{x}^{22}
-195379996286400771470\,{x}^{21} \nonumber\\
\hspace*{-\mathindent}&&+242040843303854495835\,{x}^{20}
-5651164645038662336\,{x}^{19}+11538565522879939251\,{x}^{18}
 \nonumber\\
\hspace*{-\mathindent}&&
-152433730850994094\,{x}^{17}-2210090516522360166\,{x}^{16}
+138625073940774506\,{x}^{15}
 \nonumber\\
\hspace*{-\mathindent}&&
+119612518911621073\,{x}^{14}-10912078336200028\,{x}^{13}
-3317799883792499\,{x}^{12} 
\nonumber\\
\hspace*{-\mathindent}&&
+357558807094224\,{x}^{11}+69958893047362\,{x}^{10}-1622563488732\,{x}^{9}
 \nonumber\\
\hspace*{-\mathindent}&&
-666691927048\,{x}^{8}-19283182842\,{x}^{7}+892775531\,{x}^{6}
 \nonumber\\
\hspace*{-\mathindent}&&-67944054\,{x}^{5}
-29175022\,{x}^{4}-2267238\,{x}^{3}-61803\,{x}^{2}-444\,x+9.
\eeqr
\end{footnotesize}

\vskip .1cm

\section{LCLMs of the 8th order operators
   of $\, P^{\mathrm{T}}$, $\,P^{\mathrm{I}}$, $\,P^{\mathrm{P}}$}
\label{app:LCLM}

The LCLM of the operators $ \, L_8^{\mathrm{T}}$ and $ \, L_8^{\mathrm{I}}$ 
produces a 10th order linear differential operator of the following form
\beqr
\label{LCLM8}
\hspace{-0.95in}&&   \quad  \quad \quad  \quad \quad
\mathrm{LCLM}(L_8^{\mathrm{T}},L_8^{\mathrm{I}})
 \,  \,=\,\,  \,  L_{10}^{\mathrm{TI}}  \,\,  \,=\, \,  \, 
L_3^{(1)}\cdot L_2^{(1)}\cdot L_1^{(1)}\cdot L_1^{(2)} \cdot\,  D_x^3
\nonumber  \\
\hspace{-0.95in}&&   \quad  \quad \quad  \quad \quad \quad \quad \quad
=\, \,  \,   \left(N_3\cdot N_2\cdot N_1\right) \, 
\oplus  \, L_1^{(3)} \oplus  \, L_1^{(4)} \oplus \,  L_1^{(5)}\oplus \,  D_x, 
\eeqr
where the three $L_1^{(j)}$ in the direct sum have, respectively, 
the solutions
\beqr
\label{AAA4}
\hspace{-0.95in}&&   \quad  \quad \quad \quad \quad \quad \, 
x^2, \quad \quad \quad \, 
 \frac{x}{\sqrt{1 \, -4x}} ,\quad \quad \quad \, 
 \frac{(9\,+26 \, x^2)}{\sqrt{1\,-4x}}.
\eeqr
Note that as a consequence of the direct sum structure, 
$\, L_{10}^{\mathrm{TI}}$ has
 very simple algebraic solutions which can be written
 in the following 
 form for arbitrary constants $\, A_j$
\beqr
\label{AAA4}
\hspace{-0.95in}&&   \quad  \, \,  \, \quad 
A_0 \,\, \, \,\, + A_1 \cdot \, x \,\, \, \,+A_2 \cdot \,x^2 
\,\,\,  \, + A_3 \cdot \,\frac{x}{\sqrt{1 \, -4x}} 
\,\,\, \,\, + A_4 \cdot \,\frac{(9\,+26 \, x^2)}{\sqrt{1\,-4x}}, 
\eeqr

The LCLM of the operators $ \, L_8^{\mathrm{T}}$ and $ \, L_8^{\mathrm{P}}$ 
produces a 10th order operator of the following form
\beqr
\label{LLLL4}
\hspace{-0.95in}&&   \quad  \quad\quad  \quad
\mathrm{LCLM}(L_8^{\mathrm{T}},L_8^{\mathrm{P}}) 
 \,= \, \,\,  \, L_{10}^{\mathrm{TP}}
 \, \,\,  = \,\,  \,\, 
L_3^{(2)}\cdot \, L_1^{(6)} \cdot \, L_1^{(7)} \cdot  \, 
L_2^{(2)} \cdot \, L_1^{(8)} \cdot \, L_1^{(9)}\cdot \,  L_1^{(10)}
\nonumber \\
\label{LLL4}
\hspace{-0.95in}&&   \quad  \quad \quad \quad 
 \quad \quad \quad \quad \quad
=\,\, \,\,  \left(K_3\cdot K_2\cdot N_1\right)
\oplus L_1^{(11)}\oplus L_1^{(12)}\oplus L_1^{(13)}\oplus L_1^{(14)}.
\eeqr
The $\, L_1^{(j)}$ operators in the direct sum have, respectively, 
the following solutions
\beqr
\hspace{-0.95in}&&   \quad  \,   \quad 
\frac{(3 \, -x^2)}{\sqrt{1 \, -4x}},\quad \, \, \,
 \frac{(3 \, -2x)}{\sqrt{1 \, -4x}},
\quad \,  \,\frac{(1 \,+r_1x +r_2 \, x^2)}{(1 \, -4x)},
\quad \,  \,
\frac{x \, (1 \, +r_3 \, x \, +r_4 \, x^2)}{(1 \, -4x)},
\label{L10TPsols}
\eeqr 
where the $ \, r_i$ are rational numbers with large integer 
numerators and denominators. Therefore any linear combination of 
solutions~(\ref{L10TPsols}) of $\, L_1^{(j)}$ together
 with the $\, (1-4x)$ solution of $\, N_1$ is a 
solution of $\, L_{10}^{\mathrm{TP}}$.

The LCLM of the linear differential operators $ \, L_8^{\mathrm{I}}$ 
and $ \, L_8^{\mathrm{P}}$ produces a 10th order linear differential 
operator of the following form
\beqr
\hspace{-0.95in}&&   \quad   \quad  \quad  \quad
\mathrm{LCLM}(L_8^{\mathrm{I}},L_8^{\mathrm{P}})
 \, \,=\,\, \, L_{10}^{\mathrm{IP}}
\,\, \,  =\,\, \,  \,
L_3^{(3)}\cdot L_1^{(15)}\cdot L_1^{(16)}\cdot  L_2^{(3)}\cdot L_1^{(17)}
\cdot \, L_1^{(18)} \cdot N_1
\nonumber \\
\hspace{-0.95in}&&   \quad  \quad \quad \quad \quad \quad  \quad
= \,\,\,  
\left[L_3^{(4)}\cdot \bigl((L_2^{(4)} \cdot N_1) \oplus L_1^{(19)} \bigr) \right]
\oplus L_1^{(20)}\oplus L_1^{(21)}\oplus L_1^{(22)}. 
\eeqr
The three first order operators $\,L_1^{(20)}$--$L_1^{(22)}$ have 
the following solutions
\beqr
\hspace{-0.95in}&&   \quad  \, \quad  \quad \quad \quad
\frac{(9-34x)}{\sqrt{1 \, -4x}},
\quad \quad \quad \, 
\frac{x^2}{\sqrt{1 \, -4x}},
\quad \quad \quad  \, 
s_0 \, +s_1 \, x \, +s_2 \, x^2,
\eeqr
where the $\, s_j$ are very large integers. The solution of 
$ \, L_1^{(19)}$ is of the form
\beqr
\hspace{-0.95in}&&   \quad  \quad \quad 
\quad \quad \quad \quad \quad \quad
\frac{r_0\,+r_1\,x\,+r_2\,x^2\,+r_3\,x^3}{1\,-4x}, 
\eeqr
where the $ \, r_j$ are quite large integers. 

As a consequence, any linear combination of solutions of the form
\beqr
\hspace{-0.95in}&&   \quad  \, \quad \quad \quad \quad \, 
A_0 \cdot  \, (1\, -4x)\,  \,\, \, + A_1\cdot  \, (1+\, 2x^2)
\, \,  \, + A_2\cdot \, \frac{x \cdot \, (1 \, -9x^2)}{(1 \, -4x)} \, 
\nonumber  \\
\hspace{-0.95in}&&   \quad  \quad \quad 
\quad \quad \quad  \quad \quad \,\, 
+ A_3 \cdot  \, \frac{(9-34x)}{\sqrt{1 \, -4x}} \, 
 \,  \,\, \, + A_4 \cdot  \, \frac{x^2}{\sqrt{1 \, -4x}}.
\eeqr
are solutions of $ \, L_{10}^{\mathrm{IP}}$, where we have simplified two 
of the solutions by appropriate linear combinations of the solutions 
of $\, L_1^{(19)}$ and $\, L_1^{(22)}$.

Finally, the LCLM of the three linear differential
operators $ \, L_8^{\mathrm{T}}$, 
$ \, L_8^{\mathrm{I}}$, and $ \, L_8^{\mathrm{P}}$ produces 
a 12th order operator of the following form
\beqr
\hspace{-0.98in}&&   
\mathrm{LCLM}(L_8^{\mathrm{T}},L_8^{\mathrm{I}},L_8^{\mathrm{P}}) \, 
=\, \, \,   L_{12}^{\mathrm{TIP}}  \,  \, =\, \, \,  
 L_3^{(5)}\cdot L_2^{(5)}\cdot L_1^{(23)}\cdot L_1^{(24)} \cdot
L_1^{(25)}\cdot L_1^{(26)} \cdot \,  D_x^3 
 \nonumber \\
\hspace{-0.98in}&&    \quad \, \, 
= \,  \,\,  \,  L_3^{(6)}\cdot 
\left[\left(L_2^{(6)}\cdot L_1^{(27)}\right)
\oplus L_1^{(28)} \oplus L_1^{(29)}\oplus L_1^{(30)}
 \oplus L_1^{(31)}\oplus L_1^{(32)}\oplus L_1^{(33)}  \right], 
\eeqr
which has the following seven first order operator solutions
\beqr
\hspace{-0.95in}&&   \quad  \quad \quad \, \, \, \, \, 
A_0 \, \, \, + \frac{A_1}{(1\,-4x)} \,  \,\, + A_2 \cdot \, (1\,-4x)
 \, \, \, \, + A_3\cdot \, (1\,-4x)^2 \, 
\nonumber \\
\hspace{-0.95in}&&   \quad  \quad \quad  \quad \quad \quad \quad \, \, \, 
+ \frac{A_4}{\sqrt{1\,-4x}}  \, \, \,  \,\, 
+ A_5 \cdot \,\sqrt{1\,-4x} \, \, \, \, \,+ A_6\cdot \,(1\,-4x)^{3/2}. 
\eeqr


\section{Formal power series}
\label{app:series}

The formal solutions 
of the 8th order linear operator annihilating $ \, P^{\mathrm{T}}$
have the following form at $x=0$
\beqr
\label{SSS}
\hspace{-0.95in}&&  \quad \quad \, \,
S_1 \, = \,  \,\,  1 \, -4x,  \qquad  \quad 
S_2 \, = \, \,   x^2,  \qquad  \quad 
S_3  \, = \,  \,  \frac{9-58x+26x^2}{9 \sqrt{1-4x}}, 
\nonumber  \\
\hspace{-0.95in}&&  \quad \quad \, \,
S_4  \, = \, \, \,  \sum_{n=0}^{\infty} \, c^{(4)}_n \, x^n, 
 \qquad  \quad 
S_5  \, = \,  \,  \sum_{n=-1}^{\infty} \, c^{(5)}_n \, x^n
 \,  \, + \ln(x)\cdot \, \sum_{n=0}^{\infty}c^{(4)}_n \, x^n, 
\nonumber \\
\hspace{-0.95in}&&  \quad \quad \, \,
S_6  \, = \, \, \,  \sum_{n=0}^{\infty} \, c^{(6)}_n \, x^n, 
\qquad  \quad 
S_7  \, = \,  \,  \,  \sum_{n=0}^{\infty} \, c^{(7)}_n \, x^n  \,  \,   
 \,  + \ln(x)\cdot \, \sum_{n=0}^{\infty} \, c^{(6)}_n \, x^n,
\nonumber  \\
\hspace{-0.95in}&&  \quad \quad  \, \,
S_8  \, = \,  \,\, \,   \sum_{n=0}^{\infty} \, c^{(8)}_n \, x^n \, \, 
\,  + 2\ln(x)\cdot \, \sum_{n=0}^{\infty} \, c^{(7)}_n \, x^n \, \, \, 
+ \ln^2(x) \cdot \, \sum_{n=0}^{\infty} \, c^{(6)}_n \, x^n,  
\eeqr
where both $ \, c^{(4)}_n$ and $ \, c^{(6)}_n$ are integers sequences. 
Note that only $ \, c^{(5)}_n$ starts at $\, n=\, -1$, and also note 
the factor of $ \, 2$ in the second term in $ \, S_8$. The series 
$ \, c^{(6)}_n$ is determined uniquely from the $ \, \ln^2(x)$ 
terms, and likewise, the series $ \, c^{(4)}_n$ is determined 
uniquely as the series multiplying the logarithm in the 
logarithmic solution of the operator $\, N_2 \cdot N_1$.

\vskip .1cm  

\subsection{Series at infinity}
\label{Seriesatinf}

Around $ \, y= \, 1/x \, = \, 0$, the series solutions of 
$\, L_8^{\mathrm{T}}$ 
\beqr
\label{sss}
\hspace{-0.95in}&&  \, \,
S_1 \, =\,\,  1\,-4y^{-1}, 
 \qquad \quad 
S_2 \,= \,\, y^{-2}, 
\qquad \quad 
S_3 \,= \,\, \frac{9\,-58y^{-1}\,+26y^{-2}}{\sqrt{1\,-4y^{-1}}}, 
\nonumber \\
\hspace{-0.95in}&& \, \,
S_4 \,= \,\,
\sum_{n=0}^{\infty}\,d^{(4)}_n\,y^n \, \, + \ln(x)\cdot \,(1\,-4y^{-1}),  
\qquad \,
S_5 \,= \,\, y^{-1/2} \cdot \, \sum_{n=0}^{\infty}\,d^{(5)}_n\,y^n, 
 \\
\hspace{-0.95in}&& \, \,
S_6 \,=\,\, y^{1/2} \cdot \, \sum_{n=0}^{\infty}\,d^{(6)}_n\,y^n, 
 \quad \quad
S_7 \,= \,\, y^{3/2} \cdot \, \sum_{n=0}^{\infty}\,d^{(7)}_n\,y^n, 
  \quad \quad
S_8 \,=\,\, y^{5/2} \cdot \, \sum_{n=0}^{\infty}\,d^{(8)}_n\,y^n.
\nonumber 
\eeqr


\section{Integrating $ \, N_3$ back through $\, N_2 \cdot \, N_1$}
\label{integrating}

Introducing the wronskian of $\, N_2$
\beqr
\label{wronsk}
\hspace{-0.95in}&&  \, \quad \quad   \, 
W(N_2) \,\, = \, \, \, \,
{{ 2-6\,x-60\,{x}^{2}+230\,{x}^{3}-660\,{x}^{4}
+2913\,{x}^{5}-6874\,{x}^{6} } \over {
x^3 \cdot \, (1\, -4 \, x)^4 }}, 
\eeqr
and recalling the 
solution of $\, N_1$, $\, \mathrm{Sol}(N_1) \, = \, \, 1 \, -4 \, x$, as well 
as the solutions $\, \mathrm{Sol}(N_2)$ in (\ref{N2sol}) 
or (\ref{N2solcontiguous})  of $\, N_2$
 and the solution $\, \mathrm{Sol}(N_3)$ of $\, N_3$, the 
solution $\mathrm{Sol}_3$ of (\ref{Sol3def}) entering 
into $ \, P^{\mathrm{I}}_{\mathrm{trans}}$ can be written as 
\beqr
\label{closedform}
\hspace{-0.95in}&& \quad    \,  \, \,  \,   \, 
\mathrm{Sol}(N_1) \cdot \, \Bigl( -11 \, \,\, \, 
+ \, \int  \, {{\mathrm{Sol}(N_2)} \over {\mathrm{Sol}(N_1)}} \cdot \,   \Bigl( -10 \,
 \\ 
\hspace{-0.95in}&&  \, \, \,\quad    \quad 
 + \, 90 \cdot \,  \int  \,  
\Bigl(  \int  \,  {{W} \over {\mathrm{Sol}(N_2)^2}} \cdot \, 
  \int  \, \Bigl( {{\mathrm{Sol}(N_2) \cdot \, 
\mathrm{Sol}(N_3)} \over {W(N_2) }} \cdot \, dx \Bigr) \cdot \, dx 
\Bigr) \cdot \, dx  \Bigr)  \cdot \, dx    \Bigr).
\nonumber
\eeqr


\section{Apparent singularities in $\, \mathrm{Sol}( N_3)$}
\label{app:pullback}

Let us try to understand in the solutions (\ref{solN3}) 
and (\ref{othersolN3}) the denominator 
of the pullbacks, $\, (1 -4\,{x}^{2} \, -U)$ and 
$\, (1 -4\,{x}^{2} \, +U)$, as well as the two expressions of the 
numerator of the pre-factors, namely $\, (13 \,-28\,{x}^{2}\,  -12\,U)$ 
and $\,(13 \,-28\,{x}^{2}\,  +12\,U)$. 

If one performs the resultant of these 
expressions with the definition of $\, U^2$, namely 
$\,( 1\,-12\,x^2\,-64\, x^4 \, -U^2) = \, 0$, 
one gets respectively 
$\, 4\, x^2 \cdot \,(1+\,20\,x^2)$ and  
$\, 25  \cdot \,(1+\,20\,x^2)^2$. Therefore, let us 
consider the values of $\, x$ such that $\, 1+\,20\,x^2 \, =  \, \, 0$. At 
these points one sees that  $\, \,1 -4\,{x}^{2}\,$ 
is equal to $\, +6/5$,  that   $\, U \, = \, \, \pm \, 6/5$, and that 
 $\, (13 \,-28\,{x}^{2})/12$ is  equal to $\, +6/5$. 
 
Typically, in the neighborhood of $\, 1+\,20\,x^2 \, = \, 0$, namely
for $\, x \simeq \, i/(2\cdot 5^{1/2}) \, + \, \epsilon$, we have
\begin{eqnarray}
\label{simeq2} 
\hspace*{-\mathindent}\qquad
\quad~~~ (1+\,20\,x^2)^2  &\simeq& ~ 
-80 \cdot \, \epsilon^2 \,\, +~ 160 \, i \, 5^{1/2} \, \epsilon^3
\, \, \, + \, \, \,  \cdots \\
\label{simeq}
\hspace*{-\mathindent}\qquad
(13 \,\,-28\,{x}^{2}\,  \,-12\,U)~~  &\simeq& ~
-{{625} \over {9}} \cdot \, \epsilon^2
 \,  \,\, +~{{18125} \over {162}} \, i \, 5^{1/2} \cdot \, \epsilon^3
 \, \,\,  + \, \, \,  \cdots \\
\label{simeq3}
\hspace*{-\mathindent}\qquad
\quad~ (1 -4\,{x}^{2} \, -U)^4  &\simeq& ~
\quad~ {{25} \over{81}} \cdot \, \epsilon^4 \,\, + \, \, \cdots,
\quad \,  \\
\hspace*{-\mathindent}\qquad
\quad~ {\frac {4096~{x}^{10}}{ (1 -4\,{x}^{2} \, -U)^{4}}} 
  &\simeq& ~
   -{{324} \over{78125}} \cdot \, \epsilon^{-4} \, \,+ \, \, \cdots,  
\end{eqnarray}
so that the expression
\begin{eqnarray}
\label{simeq4}
\hspace*{-\mathindent}&& \, \,\,\,
 x \cdot \, U \cdot \, \left({{13 \,-28\,{x}^{2}\,  -12\,U} \over {
 (1+ \, 20\,{x}^{2})^2 }}\right)^{1/4}
\cdot \, _2F_1\left(\left[{{1} \over {8}}, {{3} \over {8}}\right], \, [1], \, 
\, {\frac {4096\,{x}^{10}}{ (1 -4\,{x}^{2} \, -U)^{4}}}\right), 
\end{eqnarray}
behaves, up to a complex constant, like
\beqr
\label{simeq4}
\hspace{-0.95in}&&  \quad \quad 
 _2F_1\left(\left[{{1} \over {8}}, {{3} \over {8}}\right], \, [1], \, 
\,-{{324} \over{78125}} \cdot \, \epsilon^{-4}
 \, \,  + \, \, \cdots \right) 
 \, \, 
\\
\hspace{-0.95in}&&  \quad \quad \quad \quad \quad
 = \, \, \, \left({{78125} \over{324}} \cdot \, \epsilon^4 
 \, + \, \, \cdots \right) \cdot \, 
  _2F_1\left(\left[{{1} \over {8}}, {{1} \over {8}}\right], 
\, \left[ {{3} \over {4}} \right], 
\, -{{78125} \over{324}} \cdot \, \epsilon^4  \,\, + \, \, \cdots \right),
\nonumber
\eeqr
which is analytic. 

Therefore $\, 1 \, +20 \, x^2$ corresponds to an apparent singularity 
which is nevertheless necessary to write the modular form 
as a $\, _2F_1$ hypergeometric function. 
\vskip .1cm
Likewise, if one looks at $\,( 1\, -A)$, where $ \, A$ is given 
in (\ref{twopull}), it can be recast as
\beqr
\label{1-A}
\hspace{-0.95in}&&  \quad 
4 \cdot \,\left(1 -4\,{x}^{2} \, +U\right)^{-4} \, \cdot \, [(2\,x-1) 
 \, (2\,x+1) \cdot \, U \,  \, +32\,{x}^{5}+24\,{x}^{4}+10\,{x}^{2}-1]
 \nonumber\\
\hspace{-0.95in}&&  \quad 
\qquad\quad\quad\quad~\,\times \, \, 
[(2\,x-1) \, (2\,x+1) \cdot \,  U \, \, 
-32\,{x}^{5}+24\,{x}^{4}+10\,{x}^{2}-1]. 
\eeqr
The elimination of $\, U$ in the two factors 
in the numerator yield, besides $\, x= \, 0$, 
\begin{eqnarray}
\label{1-A}
\hspace{-0.95in}&& \quad \quad \quad   \,   
16\,{x}^{3}+12\,{x}^{2}+8\,x-1 \, = \, \, 0, 
\quad \, \,  \, \,  \,  \,    \,  \,  
16\,{x}^{3}-12\,{x}^{2}+8\,x+1 \, = \, \, 0.
\end{eqnarray}
Through a similar procedure as above, the roots 
of (\ref{1-A}) can also be shown to be apparent singularities.

\vskip .3cm


\section{Infinite order symmetry on the Heun function (\ref{HeunGsol})}
\label{infinite}

Let us denote $\,X \, = \,  \, x^2$. Let us consider 
the {\em genus-one} algebraic curve
\begin{eqnarray}
\label{genusone} 
\hspace{-0.95in}&& 
40960000\,{X}^{8}{Y}^{8} \cdot \, (625\,{X}^{2}+1054\,XY+625\,{Y}^{2}) 
\nonumber\\
\hspace{-0.95in}&&   \, 
+16384000\,{X}^{7}{Y}^{7}  \cdot \, (X\, +Y) \cdot \,
\left( 625\,{X}^{2}+1402\,XY+625\,{Y}^{2} \right)
\nonumber\\
\hspace{-0.95in}&&  \, \, 
 +409600\,{X}^{6}{Y}^{6} \cdot \, 
(4375\,{X}^{4}+19831\,{X}^{3}Y+31956\,{X}^{2}{Y}^{2}
+19831\,X{Y}^{3}+4375\,{Y}^{4})
\nonumber\\
\hspace{-0.95in}&&  
\,  +40960\,{X}^{5}{Y}^{5} \cdot \, (X+Y) \, \cdot \, {\cal P}_5 
\, \,  \, 
 +512\,{X}^{4}{Y}^{4} \cdot \,  {\cal P}_4 
\,  \, \,   +12800 \cdot \,{X}^{3}{Y}^{3} \cdot \,
 (X+Y)\, \cdot \, {\cal P}_3 
 \nonumber\\
\hspace{-0.95in}&& 
\,  +1600\,{X}^{2}{Y}^{2}  \cdot \, {\cal P}_2
\,  \,  +160\,X \,Y \cdot \, (X+Y) \cdot \,   {\cal P}_1
\, \, \, -200\,XY  \cdot \, (X+Y) \cdot \,   {\cal Q}_1
\, \, -50\,XY \cdot \,  {\cal R}_1
\nonumber\\
\hspace{-0.95in}&&  \, 
 +20\,XY  \cdot \, (X+Y)   \cdot \, 
({X}^{4}+2\,{X}^{3}Y+{X}^{2}{Y}^{2}+2\,X{Y}^{3}+{Y}^{4})
\nonumber\\
\hspace{-0.95in}&&  
\,  -XY \cdot \, ({X}^{4}+{X}^{3}Y+{X}^{2}{Y}^{2}
+X{Y}^{3}+{Y}^{4})
\\
\hspace{-0.95in}&& 
\, +{X}^{10}+226\,{X}^{9}Y+136451\,{X}^{8}{Y}^{2} 
-1049824\,{X}^{7}{Y}^{3}
-1268099\,{X}^{6}{Y}^{4}-1254150\,{X}^{5}{Y}^{5}
\nonumber\\
\hspace{-0.95in}&&  \, \quad \quad 
-1268099\,{X}^{4}{Y}^{6}-1049824\,{X}^{3}{Y}^{7}
+136451\,{X}^{2}{Y}^{8}+226\,X{Y}^{9} \, +{Y}^{10}
\,\,   = \, \, \,\,   0, \nonumber
\end{eqnarray}
where 
\begin{eqnarray}
\label{calP}
\hspace{-0.95in}&& \quad \quad 
{\cal P}_5 
\, \, = \, \, \, 
 4375\,\cdot \,({X}^{4}+Y^4)\,  \,  
 +777\,\cdot \, ({X}^{3}Y+X{Y}^{3})\,  \, 
  -2000 \cdot \,{X}^{2}{Y}^{2},
\nonumber\\
\hspace{-0.95in}&& \quad \quad 
{\cal P}_4 
\, \, = \, \, \,  
 21875\,\cdot \, ({X}^{6} +{Y}^{6})\,  \, 
  -99848\, \cdot \, ({X}^{5}Y +X{Y}^{5})
\nonumber\\
\hspace{-0.95in}&& \quad \quad \quad \quad \quad \quad 
 -4066848\,\cdot \, ({X}^{4}{Y}^{2}+{X}^{2}{Y}^{4})\,  
-6920598 \cdot \, {X}^{3}{Y}^{3},
\nonumber\\
\hspace{-0.95in}&& \quad \quad 
{\cal P}_3 
\, \, = \, \, \,  
 35\,\cdot \, ({X}^{6}+{Y}^{6})\,  -168\,\cdot \, ({X}^{5}Y+X{Y}^{5})
\nonumber\\
\hspace{-0.95in}&& \quad \quad \quad \quad \quad \quad 
-4876 \cdot \,({X}^{4}{Y}^{2}+{X}^{2}{Y}^{4})\, \,  
 -37398 \cdot \,{X}^{3}{Y}^{3},
\nonumber\\
\hspace{-0.95in}&& \quad \quad 
{\cal P}_2 
\, \, = \, \, \,  
7\,\cdot \, ({X}^{8}+Y^8) +133\,\cdot \, ({X}^{7}Y+X{Y}^{7})\,
 +53159\,({X}^{6}{Y}^{2}+{X}^{2}{Y}^{6})
\nonumber\\
\hspace{-0.95in}&& \quad \quad \quad \quad \quad \quad 
 +92442\,\cdot \, ({X}^{5}{Y}^{3}+{X}^{3}{Y}^{5})\,  
+102662\,{X}^{4}{Y}^{4},
 \\
\hspace{-0.95in}&& \quad \quad 
 {\cal P}_1\, \, = \, \, \,  
 {X}^{8}+ {Y}^{8}\,  +85\,\cdot \, ({X}^{7}Y+X{Y}^{7})\,  
 -42039\,\cdot \, ({X}^{6}{Y}^{2}+{X}^{2}{Y}^{6})
 \nonumber\\
\hspace{-0.95in}&& \quad \quad \quad \quad \quad \quad 
+61670\,\cdot \, ({X}^{5}{Y}^{3} +{X}^{3}{Y}^{5}) \,  
-39482 \cdot \,{X}^{4}{Y}^{4},
\nonumber\\
\hspace{-0.95in}&& \quad \quad 
 {\cal Q}_1\, \, = \, \, \,  
 2 \cdot \,({X}^{6}+Y^6) \,  -143 \cdot \,({X}^{5}Y+X{Y}^{5}) \,  \,  
+254 \cdot \,({X}^{4}{Y}^{2}+{X}^{2}{Y}^{4})
 \,  \,  -149\,{X}^{3}{Y}^{3},
\nonumber\\
\hspace{-0.95in}&& \quad \quad 
 {\cal R}_1\, \, = \, \, \,  
{X}^{6}+{Y}^{6} \, 
-70\,\cdot \, ({X}^{5}Y +X{Y}^{5})\,  \,  
 -60\,\cdot \,({X}^{4}{Y}^{2} +{X}^{2}{Y}^{4}) \,  \,  -60\,{X}^{3}{Y}^{3}.
\nonumber
\end{eqnarray}

One can write $\, Y$ in (\ref{genusone}) as 
a series expansion in $\, X$, namely
\begin{eqnarray}
\label{seriesYfirst}
\hspace{-0.95in}&&
Y \, \,= \, \,\,\, {X}^{5} \,  +20\,{X}^{6} +350\,{X}^{7} +5600\,{X}^{8}
+86725\,{X}^{9} \,  +1319200\,{X}^{10}
\nonumber\\
\hspace{-0.95in}&&  \, \quad
+19894850\,{X}^{11} \, +298777600\,{X}^{12} \, 
+4479731850\,{X}^{13} \, 
+67155693600\,{X}^{14}
\nonumber\\
\hspace{-0.95in}&&  \, \quad
+1007421693450\,{X}^{15}
+15130465630600\,{X}^{16}
\nonumber\\
\hspace{-0.95in}&&  \, \quad
+227576601943225\,{X}^{17} \, 
+3428478377045600\,{X}^{18} \, 
+51737085633726100\,{X}^{19}
\nonumber\\
\hspace{-0.95in}&&  \, \quad
+782050723102305200\,{X}^{20} \, \, 
+11841094422935733850\,{X}^{21}
\nonumber\\
\hspace{-0.95in}&&  \, \quad
+179579006419196877600\,{X}^{22} \, \, 
+2727744732078726781850\,{X}^{23}
\nonumber\\
\hspace{-0.95in}&&  \, \quad
+41496463049656511818600\,{X}^{24} \, \, 
+632194923237727485070075\,{X}^{25}
\nonumber\\
\hspace{-0.95in}&&  \, \quad
+9644872198249006185042100\,{X}^{26} \, \, 
+147340024316081333011633850\,{X}^{27}
\nonumber\\
\hspace{-0.95in}&&  \, \quad
+2253708185187840469204115600\,{X}^{28} \, \, 
+34514542785442406208079674225\,{X}^{29}
\nonumber
\end{eqnarray}
We have the following identity on the pullbacks 
\begin{eqnarray}
\label{identpull}
\hspace{-0.95in}&& \quad \quad \quad 
 {\frac {4096\,{X}^{5}}{ \left[1 -4\,X \,
 + \, \sqrt{(1\,  -16 \, X)  \, (1\, +4\,X)}\right]^{4}}}
\nonumber\\
\hspace{-0.95in}&&  \, \quad \quad \quad \quad \quad  \quad \quad
\, \, = \, \, \,  \,   
 {\frac {4096\,{Y}^{5}}{ 
\left[1 -4\,Y \, - \, \sqrt{(1\,  -16 \, Y)  \, (1\, +4\,Y)}\right]^{4}}}
 \\
\hspace{-0.95in}&&  \, \quad \quad \quad \quad 
\, \, = \, \, \,   \,  \,   
256\,{X}^{5}\,  \,   +5120\,{X}^{6}\, \, 
  +89600\,{X}^{7}\,  +1433600\,{X}^{8}\,  
+22201600\,{X}^{9}\, 
\nonumber\\
\hspace{-0.95in}&&  \, \quad \quad \quad  \quad \quad \quad \quad \quad \, 
+337689600\,{X}^{10} \, 
+5092057600\,{X}^{11}\,   \, 
 +76458905600\,{X}^{12}\,  
\,\, \,    + \, \, \cdots
 \nonumber
\end{eqnarray}
from which one deduces from (\ref{solN3}), the following 
{\em infinite order automorphism  identity on a Heun function}
\begin{eqnarray}
\label{identHeunG}
\hspace{-0.95in}&& \quad \quad \quad \quad \, \, \,
 {\cal A}_1(X) \cdot \, \mathrm{Heun}\Bigl( -{{1} \over {4}}, \, {{1} \over {16}}, \, \,
 {{3} \over {8}}, \, {{5} \over {8}},\, 1, \, {{1} \over {2}}, \,  -4 \, X\Bigr)
\nonumber\\
\hspace{-0.95in}&&  \, \quad \quad \quad \quad \quad \quad \, \, \,
= \, \, \, \,    {\cal A}_2(Y) \cdot \,
\mathrm{Heun}\Bigl( -{{1} \over {4}}, \, {{1} \over {16}}, \, \,
 {{3} \over {8}}, \, {{5} \over {8}},\, 1, \, {{1} \over {2}}, \,  -4 \, Y\Bigr), 
\end{eqnarray}
where 
\begin{eqnarray}
\label{identHeunG}
\hspace{-0.95in}&&  \, 
 {\cal A}_1(X) \, \, = \, \, \,
\left[{{ (1+20\, X)^2 \cdot \, (1 \, -12 \, X  \, -64 \, X^2)} \over {
(1\,  -16 \, X)^2  \cdot \,  (1\, +4\,X)^2  \cdot \,  
[13 \,-28\, X\,  -12\,\sqrt{(1\,  -16 \, X)  \, (1\, +4\,X)}]
}}\right]^{1/4}, 
\nonumber\\
\hspace{-0.95in}&&  \, 
  {\cal A}_2(Y) \, \,  = \, \, 
\left[{{ 25 \cdot \, (1+20\, Y)^2 \cdot \, 
(1 \, -12 \, Y  \, -64 \, Y^2)} \over {
(1\,  -16 \, Y)^2  \cdot \,  (1\, +4\,Y)^2  \cdot \,  
[13 \,-28\,Y \,  +12\,\sqrt{(1\,  -16 \, Y)  \, (1\, +4\,Y)}]
}}\right]^{1/4}.
\nonumber
\end{eqnarray}

\vskip .5cm 

\pagebreak

\end{document}